\documentclass[12pt]{article}
\usepackage{url}
\input{epsf}

\font\bm=cmmib10 at 10pt
\font\bms=cmmib10 at 7pt \textfont9=\bm \scriptfont9=\bms
\usepackage{amsfonts}
\usepackage{multirow}
\mathchardef\balpha= "790B
\mathchardef\bbeta= "790C
\mathchardef\bTheta= "7902
\mathchardef\bzeta= "7910
\mathchardef\bOmega= "790A
\mathchardef\bGamma= "7900
\mathchardef\bDelta= "7901
\mathchardef\bPhi= "7908
\mathchardef\bphi= "791E
\mathchardef\bomega= "7921
\mathchardef\bxi= "7918
\mathchardef\bet= "7911
\mathchardef\brho= "791A
\mathchardef\btau= "791C
\mathchardef\bmu= "7916
\mathchardef\bvarpi= "7924

\def \lvec{(\kern-.26em(}
\def\pmb#1{\setbox0=\hbox{#1}%
\def \lvec{(\kern-.26em(}
\kern-.025em\copy0\kern-\wd0
\kern.05em\copy0\kern-\wd0
\kern-.025em\raise.0433em\box0 }
\mathchardef\btheta= "7912

\usepackage{amsmath}
\usepackage{amssymb}
\usepackage{authblk}
\usepackage{url}
\providecommand{\keywords}[1]{\textbf{\textit{Keywords:  }} #1}
\usepackage{graphicx}% Include figure files
\usepackage{dcolumn}% Align table columns on decimal point

\begin{document}
\title{Effect of Greenhouse Gases on Thermal Emissivity by Clouds}
\author[1]{ W. A. van Wijngaarden}
\author[2] {W. Happer}
\affil[1]{Department of Physics and Astronomy, York University, Canada}
\affil[2]{Department of Physics, Princeton University, USA}
\renewcommand\Affilfont{\itshape\small}
\date{\today}
\maketitle
\begin{abstract}
\noindent Greenhouse gases, most importantly water vapor,  increase the emissivity and decrease the albedo of clouds for thermal radiation. The modifications, which can be of order 10\% for optically thick clouds, depend on the attenuation coefficient $\alpha^{\{g\}}$ of the greenhouse gases, and also  on the attenuation coefficient, $\alpha^{\{c\}}$, the single-scattering albedo $\tilde \omega^{\{c\}}$,  and the scattering phase function $p^{\{c\}}(\mu,\mu')$ of the cloud particulates. Cold, high-altitude clouds with low partial pressures of water vapor have smaller emissivities for thermal radiation and larger albedos than otherwise identical but warmer low-altitude clouds with higher partial pressures of water vapor. In $2n$-stream scattering theory, these phenomena can be quantified with the intensity emissivities $\varepsilon_{i}$ of the streams $i=1,2,3,\ldots, 2n$, and with upward or downward flux emissivities, $\varepsilon_{\bf u}$ and $\varepsilon_{\bf d}$. The emissivities are the ratios of the outgoing thermal intensities or fluxes  to those  of a reference black cloud. Emission from optically-thick, isothermal clouds with scattering, as well as absorption and emission, is limb darkened. Intensity emissivities $\varepsilon_i$  for streams that are nearly normal to the  cloud surface are larger than those of more nearly horizontal streams.  The limb darkening increases with increasing values of the single scattering albedo $\tilde \omega$. 
For fixed values of $\tilde\omega$, the onset of limb darkening with increasing zenith angle is more abrupt for phase functions with more forward scattering. Black clouds,  which have only absorption and emission but no scattering, have unit (Lambertian) emissivities, $\varepsilon_i = 1$, for all stream directions.
\end{abstract}

\newpage
\tableofcontents
\newpage
\section{Introduction}
Greenhouse gases in clouds of water droplets, ice crystallites, and other particulates  increase the emissivity and decrease the albedo of the clouds for thermal radiation. 
In the air between cloud particulates, greenhouse-gas molecules emit and absorb radiation very efficiently at their resonance frequencies but have negligible scattering.
Some of the molecular emission can be absorbed by cloud particulates or by other molecules, but single or multiple  scattering by the particulates releases the remainder through the top and bottom surfaces of the cloud and increases the thermal emissivity, in comparison to a cloud with no greenhouse gases. The greenhouse gases also increase the probability that an external photon, incident onto the top or bottom of the cloud, will be absorbed before it can be scattered back out. This decreases the albedo of the cloud.

The steady-state equation of radiation transfer  for axially symmetric conditions in clouds is 
\begin{equation}
\left(\frac{\mu}{\alpha}\frac{\partial}{\partial z}+1\right) I(\mu, z) =(1-\tilde \omega) B(z)+\frac{\tilde\omega}{2}
\int_{-1}^1 d\mu' p(\mu,\mu')I(\mu',z).\label{rtc2}
\end{equation}
Here $I(\mu, z)=I(\mu, z,\nu)$ is the monochromatic radiation {\it intensity} of frequency $\nu$ at the altitude $z$,  propagating with a direction cosine $\mu =\cos\theta$ with respect to the vertical. The zenith angle of the intensity is $\theta$. The {\it attenuation coefficient}, $\alpha=\alpha(z,\nu)$, gives the infinitesimal probability $\partial P=\alpha\,\partial z$ that a photon is absorbed or scattered after propagating an infinitesimal distance $\partial z$.

On the right of (\ref{rtc2}), $\tilde\omega$ denotes the {\it single-scattering albedo}, the probability that a photon colliding with a cloud particulate or greenhouse-gas molecule is scattered, rather than absorbed and converted to heat.    The monochromatic {\it Planck intensity} 
$B=B(z)= B(T,\nu)$ of spatial frequency $\nu$ and absolute temperature $T=T(z)$ is
\begin{equation}
B=\frac{2h_{\rm P}c^2\nu^3}{e^{\nu c\, h_{\rm P}/(k_{\rm B}T)}-1}.
\label{rtc4}
\end{equation}
Here $h_{\rm P}$ is Planck's constant, $c$ is the speed of light and $k_{\rm B}$ is Boltzmann's constant. The kernel of the integral transform on the right of (\ref{rtc2}) is the scattering phase function $p(\mu,\mu')$. The probability $dP$ that a photon propagating with the initial direction cosine $\mu'$ is scattered into a new direction cosine between $\mu$ and $\mu+d\mu$ is $dP= p(\mu,\mu')d\mu/2$.
The well-known integro-differential equation (\ref{rtc2}) is the same as Eq, (1) in the recent paper {\it Radiation Transfer in Cloud Layers}\cite{WH4}, to which we will make frequent reference.

The integro-differential equation (\ref{rtc2}) is difficult to solve efficiently and accurately enough to model radiation transfer in clouds. But very nearly the same results can be obtained with the much simpler $2n$-stream approximation to (\ref{rtc2}), given by Eq. (55) of reference\,\cite{WH4},
\begin{equation}
\frac{d}{d\tau}|I\}=\hat \kappa\bigg( |B\}- |I\} \bigg).
\label{rtc6}
\end{equation}
This is an ordinary, linear differential equation for the $2n\times 1$ radiation intensity vector $|I\}$ with a $2n\times 1$ thermal source vector $|B\}$ and a $2n\times 2n$ exponentiation rate operator $\hat\kappa$. The  increment of optical depth  is
\begin{equation}
d\tau=\alpha dz.
\label{rtc8}
\end{equation}
A more detailed discussion of  $|I\}$, $|B\}$ and $\hat \kappa$, 
and how to efficiently solve (\ref{rtc6}) can be found  in reference\,\cite{WH4}. The $2n$-stream method of analyzing radiative transfer is a generalizaion of the $2$-stream method described in Schuster's paper, {\it Radiation Through a Foggy Atmosphere} \,\cite{Schuster}, published in  1905.  Using $2n$ streams along the Gauss-Legendre sample directions to generalize Schuster's work was first suggested by Wick in his paper 
{\it  \"Uber ebene Diffusionsprobleme}\,\cite{Wick}, published in 1943.

\section{Radiation Transfer Parameters \label{rtp} }
In this section, we  discuss the basic parameters of the equation of radiative transfer (\ref{rtc2}): the attenuation coefficient $\alpha$, the single-scattering albedo $\tilde\omega$ and the scattering phase function $p(\mu,\mu')$.

\subsection{The attenuation coefficient $\alpha$}
The attenuation coefficient $\alpha$ of (\ref{rtc2}) is the sum of a part $\alpha^{\{g\}}$ due to greenhouse gases and a part $\alpha^{\{c\}}$ due to cloud particulates.
\begin{equation}
\alpha =\alpha^{\{g\}}+\alpha^{\{c\}}.
\label{df2}
\end{equation}
In accordance with (\ref{rtc8}),
the attenuation coefficient (\ref{df2}) can be used to write the optical depth $\tau=\tau(z)$ at the altitude $z$ above the bottom of a cloud as

\begin{equation}
\tau = \int_0^z dz' \alpha(z') .
\label{df8d}
\end{equation}

In Eq. (25) of reference\,\cite{WH1}, we showed that the spatial attenuation coefficient $\alpha^{\{g\}}=\alpha^{\{g\}}(z,\nu)$ of radiation of frequency $\nu$  at the altitude $z$ (denoted by  $\kappa(z,\nu)$ in reference\,\cite{WH1}) is almost purely absorptive (there is negligible scattering of thermal radiation by greenhouse gases) and can be written as 
\begin{equation}
\alpha^{\{g\}}(z,\nu)=\sum_i N^{\{i\}}(z)\sigma^{\{i\}}(z,\nu),
\label{df6}
\end{equation}
where $N^{\{i\}}(z)$ is the number density of greenhouse-gas molecules of type $i$, with $i=$ H$_2$O, CO$_2$, O$_3$, etc. 
As shown in Eq. (3) of reference\,\cite{WH1},
the absorption (and  stimulated emission) cross section $\sigma^{\{i\}}(z,\nu)$  at the altitude $z$ is normally assumed to be the  sum of partial cross sections $\sigma_{ul}^{\{i\}}(z,\nu)$, 
\begin{equation}
\sigma^{\{i\}}(z,\nu)=\sum_{ul} \sigma_{ul}^{\{i\}}(z,\nu),
\label{lbl16}
\end{equation}
corresponding to the Bohr spatial frequencies,
\begin{equation}
\nu_{ul}^{\{i\}}= \frac{E_u^{\{i\}}-E_l^{\{i\}}}{h_{\rm P}c}
\label{lbl2}
\end{equation}
for transitions between an upper ($u$) vibration-rotation level of energy $E_u^{\{i\}}$ and a lower level of energy $E_l^{\{i\}}$. The absorption cross sections $\sigma^{\{i\}}(z,\nu)$ of (\ref{lbl16}) depend strongly on the spatial frequency $\nu$. The absorption cross sections $\sigma^{\{i\}}(z,\nu)$  also depend on the altitude $z$. This is because the fractions of molecules in the lower level $l$ and upper level $u$ of the molecule are determined by the altitude-dependent temperature  $T(z)$. Molecules in the lower level attenuate radiation when they absorb photons and transition to the upper level. Molecules in the upper level amplify radiation when they are stimulated to  emit photons and transition to the lower level.  The widths of the individual vibration-rotation absorption lines are functions of the altitude-dependent  pressure $p(z)$ and temperature $T(z)$, and this also contributes to the altitude dependence of the cross sections.

As shown in Eq. (4) of reference\,\cite{WH1},
the partial cross section $\sigma_{ul}^{\{i\}}$ of (\ref{lbl16}) is normally written as  the product of 
a line intensity, $S_{ul}^{\{i\}}=S_{ul}^{\{i\}}(T)$, which depends on  temperature $T$, and
a  line-shape function, $ G_{ul}^{\{i\}}= G_{ul}^{\{i\}}(\nu, p,T)$, which depends on  the frequency $\nu$, the pressure $p$ and the temperature $T$.
\begin{equation}
\sigma_{ul}^{\{i\}}=S_{ul}^{\{i\}} G_{ul}^{\{i\}}.
\label{lbl18}
\end{equation}
The  line shape functions, $ G_{ul}^{\{i\}}$,  are normalized to have unit area,
\begin{equation}
\int_0^{\infty} G_{ul}^{\{i\}}d\nu=1.
\label{lbl20}
\end{equation}
Since we use spatial frequencies $\nu$, with units of waves per centimeter or cm$^{-1}$, the units of $ G_{ul}^{\{i\}}$ are cm.  As discussed in Section {\bf 3.1} of reference\,\cite{WH1}, accurate and complete values of the line intensities $S_{ul}^{\{i\}}$ for the most important greenhouse gases can be obtained from the HITRAN data base\,\cite{HITRAN}.

The jagged green line of Fig \ref{emis3} shows the attenuation coefficient  $\alpha^{\{g\}}(\nu)$ of (\ref{df6}) for thermal-radiation frequencies in the range 500 to 700 cm$^{-1}$. The values were calculated at a sea level pressure of 1000 mb, at 100\% relative humidity.  At the Earth's average surface temperature of 288.7 K, the saturation vapor pressure of liquid water is $p^{\{\rm sat\}}= 17.7$ hPa. Absorption by pure rotational transitions of H$_2$O molecules is responsible for most of the attenuation for frequencies $\nu<620$ cm$^{-1}$. The air has 400 ppm of CO$_2$, which is responsible for the absorption band centered at the CO$_2$ bending mode frequency $\nu = 667$ cm$^{-1}$. Nitrous oxide N$_2$O at a concentration of 0.32 ppm makes a very small contribution to $\alpha^{\{g\}}$ for frequencies near its bending-mode frequency $\nu = 588$ cm$^{-1}$. The other naturally occuring greenhouse gases, ozone, O$_3$,  and methane, CH$_4$, make negligible contributions for the frequency range of Fig. \ref{emis3}. For the range of frequencies $\nu$ shown in Fig. \ref{emis3}, the magnitude of $\alpha^{\{g\}}$ varies in a complicated way, from  a minimum close to $10^{-5}$ m$^{-1}$ to nearly 10 m$^{-1}$, about six orders of magnitude. 

Cloud particulates also attenuate radiation, but unlike greenhouse gases, where essentially all of the attenuation is due to absorption (with negligible scattering), the attenuation by cloud particulates is partly due to absorption and partly due to scattering.
The cloud attenuation rate $\alpha^{\{c\}}$ varies much more slowly with frequency $\nu$ than the attenuation rate $\alpha^{\{g\}}$ due to greenhouse gases. To simplify subsequent discussions,  we have assumed a frequency-independent cloud attenuation rate from clouds,
\begin{equation}
\alpha^{\{c\}} =0.01 \hbox{ m}^{-1} ,
\label{df8b}
\end{equation}
 and indicated it as the dashed red line in Fig. \ref{emis3}. This corresponds to a thermal radiation ``visibility'' of $1/\alpha^{\{c\}} =100 $ m through the cloud particulates.

\subsection{The single-scattering albedo $\tilde\omega$}
\begin{figure}[t]
\includegraphics[height=80mm,width=1\columnwidth]{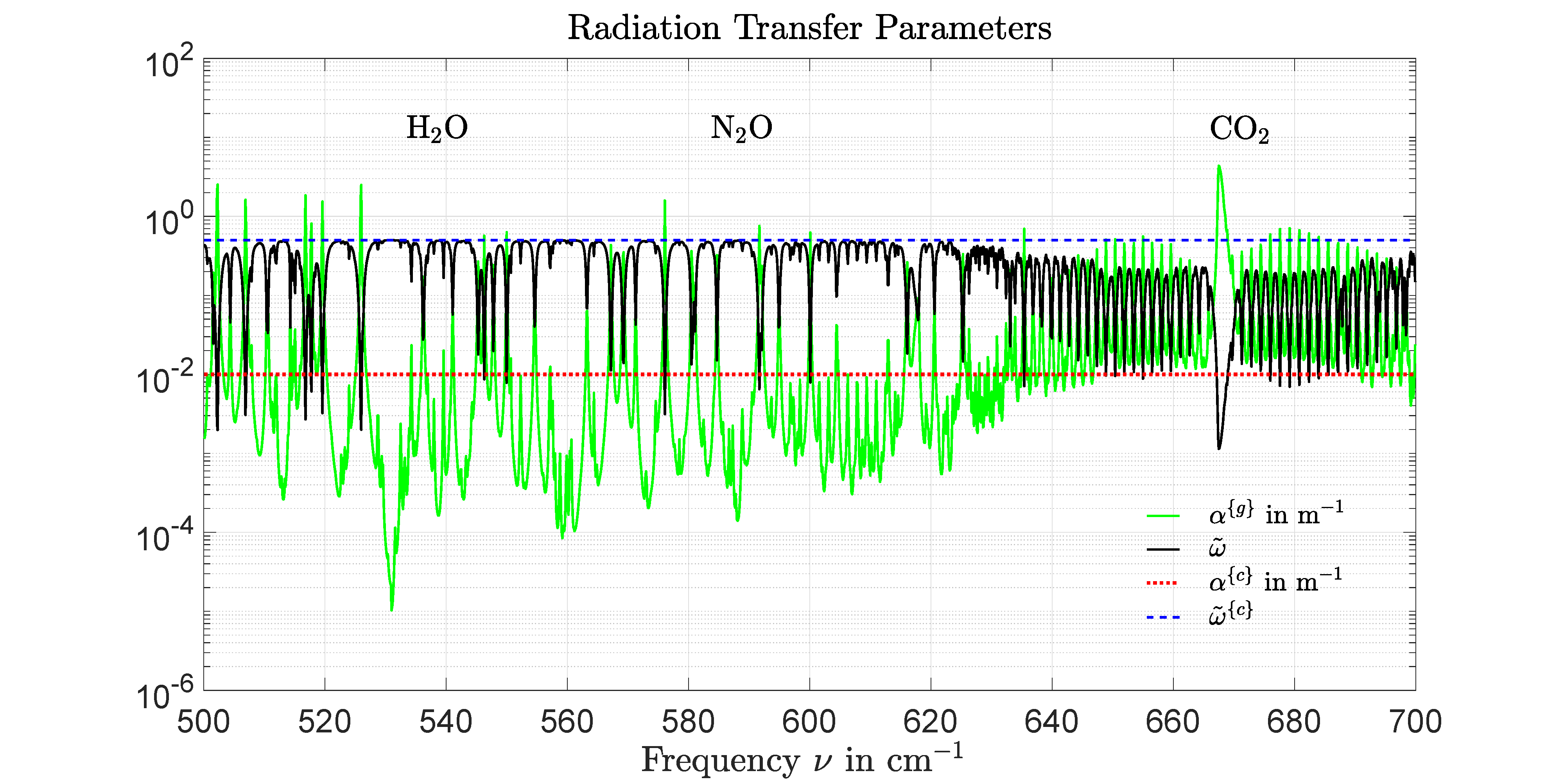}
\caption {The jagged green line is the  attenuation coefficient $\alpha^{\{g\}}$ of (\ref{df6}) at sea level due to greenhouse gases. The complicated dependence on frequency comes from
the line structure of the greenhouse-gas molecules. Simplified parameters for cloud particulates are the dashed red line,  showing the frequency-independent  attenuation  coefficient $\alpha^{\{c\}}=0.01$ m$^{-1}$ of (\ref{df8b}) and the  dashed cyan line, showing the frequency-independent single-scattering albedo $\tilde\omega^{\{c\}}=0.5$  of (\ref{df12d}). The jagged black line is the overall single scattering albedo $\tilde \omega $ of (\ref{df12}). The symbols H$_2$O and CO$_2$ mark frequency ranges where absorption by those molecules predominates. There is also a very small contribution from N$_2$O close to 588 cm$^{-1}$.}
\label{emis3}
\end{figure}

In (\ref{df2}),  the attenuation coefficient $\alpha^{\{c\}}$  due to cloud particulates consists of two parts,
\begin{equation}
\alpha^{\{c\}} =\alpha^{\{c\}}_a+\alpha^{\{c\}}_s.
\label{df8}
\end{equation}
The part $\alpha^{\{c\}}_{a}$ describes the rate at which radiation is absorbed and converted to heat; $\alpha^{\{c\}}_{s}$  describes the rate at which radiation is scattered to other directions with no conversion to heat.  The attenuation coefficient $\alpha^{\{g\}}$ due to greenhouse gases can be written in an analogous way
\begin{equation}
\alpha^{\{g\}} =\alpha^{\{g\}}_a+\alpha^{\{g\}}_s.
\label{df8a}
\end{equation}
For the conditions of Earth's atmosphere, the scattering of thermal radiation by greenhouse-gas molecules is completely negligible compared to absorption, as discussed in connection with Eq. (30) of reference\,\cite{WH0}. It is therefore an excellent approximation to set
\begin{equation}
\alpha^{\{g\}}_s=0\quad\hbox{and}\quad \alpha^{\{g\}}_a=\alpha^{\{g\}}.
\label{df8c}
\end{equation}
We define the single-scattering albedos of cloud particulates and greenhouse gases as
\begin{equation}
\tilde\omega^{\{c\}} =\frac{\alpha^{\{c\}}_s}{\alpha^{\{c\}}}\quad\hbox{and}\quad
\tilde\omega^{\{g\}} =\frac{\alpha^{\{g\}}_s}{\alpha^{\{g\}}}.
\label{df10}
\end{equation}
The single-scattering albedo $\omega^{\{c\}}$ of clouds is poorly known for thermal radiation. It  probably depends more on frequency than the cloud attenuation rate $\alpha^{\{c\}}$, which we took to be independent of frequency in (\ref{df8b}). But for illustrative purposes, we will assume that the single-scattering albedo for cloud particulates is also frequency-independent and has the value, 
\begin{equation}
\tilde \omega^{\{c\}}=0.5.
\label{df12d}
\end{equation}
The dashed cyan line of Fig. \ref{emis3} shows the assumed value of $\omega^{\{c\}}$ from (\ref{df12d}).
The approximation (\ref{df8c}) corresponds to a vanishing single-scattering albedo for greenhouse gases,
\begin{equation}
\tilde \omega^{\{g\}}=0.
\label{df12a}
\end{equation}
The overall single scattering albedo is
\begin{equation}
\tilde\omega =\frac{\alpha^{\{c\}}_s+\alpha^{\{g\}}_s}{\alpha^{\{c\}}+\alpha^{\{g\}}}=\frac{\alpha^{\{c\}}\tilde \omega^{\{c\}}+\alpha^{\{g\}}\tilde\omega^{\{g\}}}{\alpha}.
\label{df12}
\end{equation}
For a cloud containing greenhouse gases with vanishing single-scattering albedos, $\tilde \omega^{\{g\}}=0$,  the overall single-scattering albedo (\ref{df12}) becomes
\begin{equation}
\tilde\omega =\frac{\alpha^{\{c\}}\tilde \omega^{\{c\}}}{\alpha}.
\label{df12b}
\end{equation}
We have plotted  $\tilde \omega$ of (\ref{df12b}) as the jagged black line in Fig. \ref{emis3}.  The large changes of $\tilde\omega$ with frequency $\nu$ are due to the large changes of the greenhouse-gas attenuation rate $\alpha^{\{g\}}$. For atmospheric ``window'' frequencies where $\alpha^{\{g\}}\ll\alpha^{\{c\}}$, the overall  single-scattering albedo is nearly the same as that of cloud particlates, $\tilde\omega\to\tilde\omega^{\{c\}}$. Though not shown in Fig. \ref{emis3}, nearly all of the frequencies between 800 to 1000 cm$^{-1}$ are window frequencies. Near the centers of greenhouse-gas absorption lines, where  $\alpha^{\{g\}}\gg \alpha^{\{c\}}$, the overall  single-scattering albedo is nearly zero
 $\tilde\omega\to\ 0$.

\subsection{The scattering phase function $p(\mu,\mu')$}
The probability that a photon with an initial direction cosine $\mu'$ is scattered by a cloud particulate into direction cosines between $\mu$ and $\mu+d\mu$ is  $dP = d\mu\, p^{\{c\}}(\mu,\mu')/2$.  The scattering phase function   $p^{\{g\}}(\mu,\mu')$ for greenhouse gases has an analogous meaning.
One can use arguments like those leading to (\ref{df12}) to write the phase function $p(\mu,\mu')$ that appears in (\ref{rtc2}), which accounts for both cloud particulates and greenhouse gases, as 
\begin{equation}
p(\mu,\mu')= \frac{\alpha^{\{c\}}\tilde \omega^{\{c\}}p^{\{c\}}(\mu,\mu')+\alpha^{\{g\}}\tilde\omega^{\{g\}}p^{\{g\}}(\mu,\mu')}{\alpha\tilde\omega}.
\label{spf0a}
\end{equation}
For the special case (\ref{df12a}) that $\tilde\omega^{\{g\}}=0$, the phase function (\ref{spf0a}) reduces to
\begin{equation}
p(\mu,\mu')=p^{\{c\}}(\mu,\mu').
\label{spf0b}
\end{equation}
Greenhouse gases between the cloud particulates make no difference to the scattering phase function $p(\mu,\mu')$, which is determined entirely by cloud particulates. But for frequencies strongly absorbed by greenhouse gases, the single scattering albedo $\tilde\omega$ can be reduced to 1\% or less, as shown in Fig. \ref{emis3}.

We assume rotational symmetry for the scattering of radiation by cloud particulates. This is  a good approximation for most clouds of liquid water droplets, which are nearly spherical.  For clouds of ice crystallites, there will be rotational symmetry if non-spherical crystallites have random orientations. The assumption of random orientations can be violated, for example, for rare winter conditions when  {\it light pillars}\,\cite{light-pillars} form from plate-like, hexagonal crystallites, suspended with a horizontal orientation in calm air. We will not consider such unusual situations in this paper.
Rotational symmetry allows us to write the scattering phase function of (\ref{rtc2}) in the form
\begin{equation}
p(\mu,\mu')=\sum_{l=0}^{2m-1} p_l(2l+1)P_l(\mu)P_l(\mu'),
\label{spf2}
\end{equation}
as given by Eq. (126) of reference\,\cite{WH1}.  Here  $P_l(\mu)$ and $P_l(\mu')$ are Legendre polynomials. The scattering phase function $p(\mu,\mu')$ is nonnegative, and is invariant to the exchange of $\mu$ and $\mu'$
\begin{equation}
 p(\mu,\mu')=p(\mu',\mu)\ge 0.
\label{spf4}
\end{equation}
In keeping with its significance as a probability density,  the phase function satisfies the identity
\begin{equation}
\frac{1}{2}\int_{-1}^1 d\mu \,p(\mu,\mu')=1.
\label{spf6}
\end{equation}
The possible values of the multipole coefficients $p_l$ of (\ref{spf2}) are constrained by (\ref{spf4}) and (\ref{spf6}).

\begin{table}[t]
\begin{center}
\begin{tabular}{|c|c|c|c|c|}
 \hline
 & Isotropic & Rayleigh & Forward & Backward \\ [0.5ex]
 \hline\hline
$p_{0}$  & 1 & 1 & 1 &1\\
 \hline
$p_{1}$&0&0&.8182&-.8182\\
 \hline
$p_{2}$&0&0.1&.7273&.7273\\
 \hline
$p_{3}$&0&0&.5967&-.5967\\
 \hline
$p_{4}$&0&0&.4988&.4988\\
 \hline
 $p_{5}$&0&0&.3869&-.3869\\
 \hline
 $p_{6}$&0&0&.2937&.2937\\
 \hline
 $p_{7}$&0&0&.2016&-.2016\\
 \hline
 $p_{8}$&0&0&.1209&.1209\\
 \hline
 $p_{9}$&0&0&.0573&-.0573\\
 \hline
 \hline
\end{tabular}
\end{center}
\caption{ Numerical values of the multipole coefficients  $p_l$ for the phase functions  of (\ref{spf2}) or (\ref{spf10}). The numbers  $p_l=\varpi^{\{m\}}_l$ in the fourth column and  $p_l=(-1)^l\varpi^{\{m\}}_l$ in  the  fifth column give phase functions  that can be constructed from the first $2m=10$ Legendre polynomials, and have the maximum possible forward and backward values. The values of  $\varpi^{\{m\}}_l$ are from Table 1 of reference\,\cite{WH1}, with $m=5$.
\label{pl}}
\end{table}

We will consider four types of scattering, isotropic, Rayleigh, forward and backward. Numerical values of $p_l$  for these scattering types are listed in Table \ref{pl}.  The forward scattering phase function, $p(\mu,\mu') =\varpi^{\{m\}}(\mu,\mu')$, is given by Eq. (141) of reference \cite{WH1}, and is constructed from the first $2m=10$ Legendre polynomials to give the maximum possible forward scattering, $\varpi^{\{m\}}(1,1)=m(m+1)=30$.  The backward scattering phase function is $p(\mu,\mu') =\varpi^{\{m\}}(-\mu,\mu')$.

\begin{figure}[t]
\includegraphics[height=80mm,width=1\columnwidth]{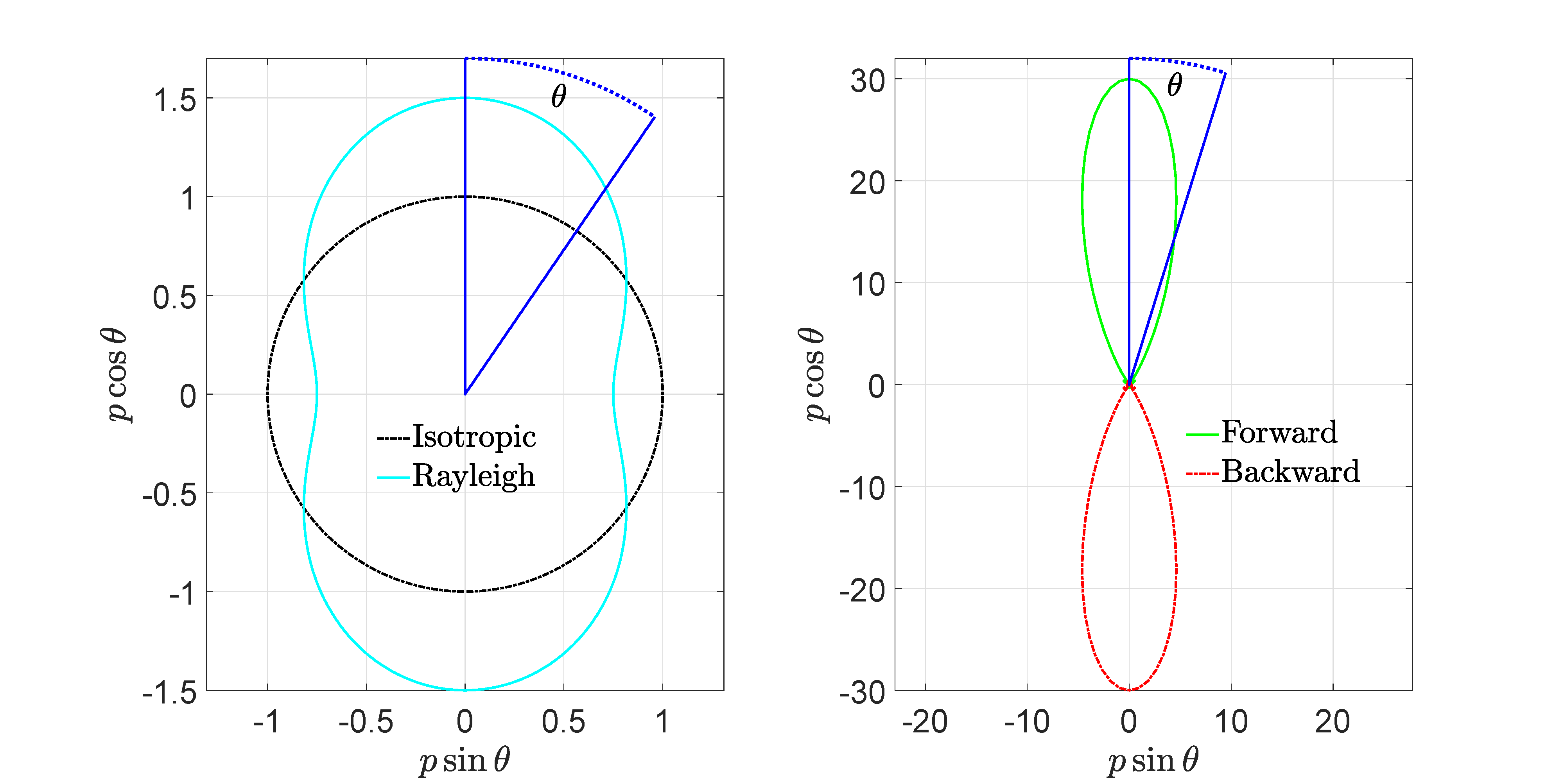}
\caption {Polar plots of the scattering phase functions $p=p(\mu)$ of (\ref{spf10}), evaluated with the multipole coefficients of Table \ref{pl}. The radiation is scattered from the initial vertical direction, with direction cosine $\mu'=1$ into the angle $\theta=\cos^{-1}\mu$.}
\label{phase1}
\end{figure}

\begin{figure}[t]
\includegraphics[height=80mm,width=1\columnwidth]{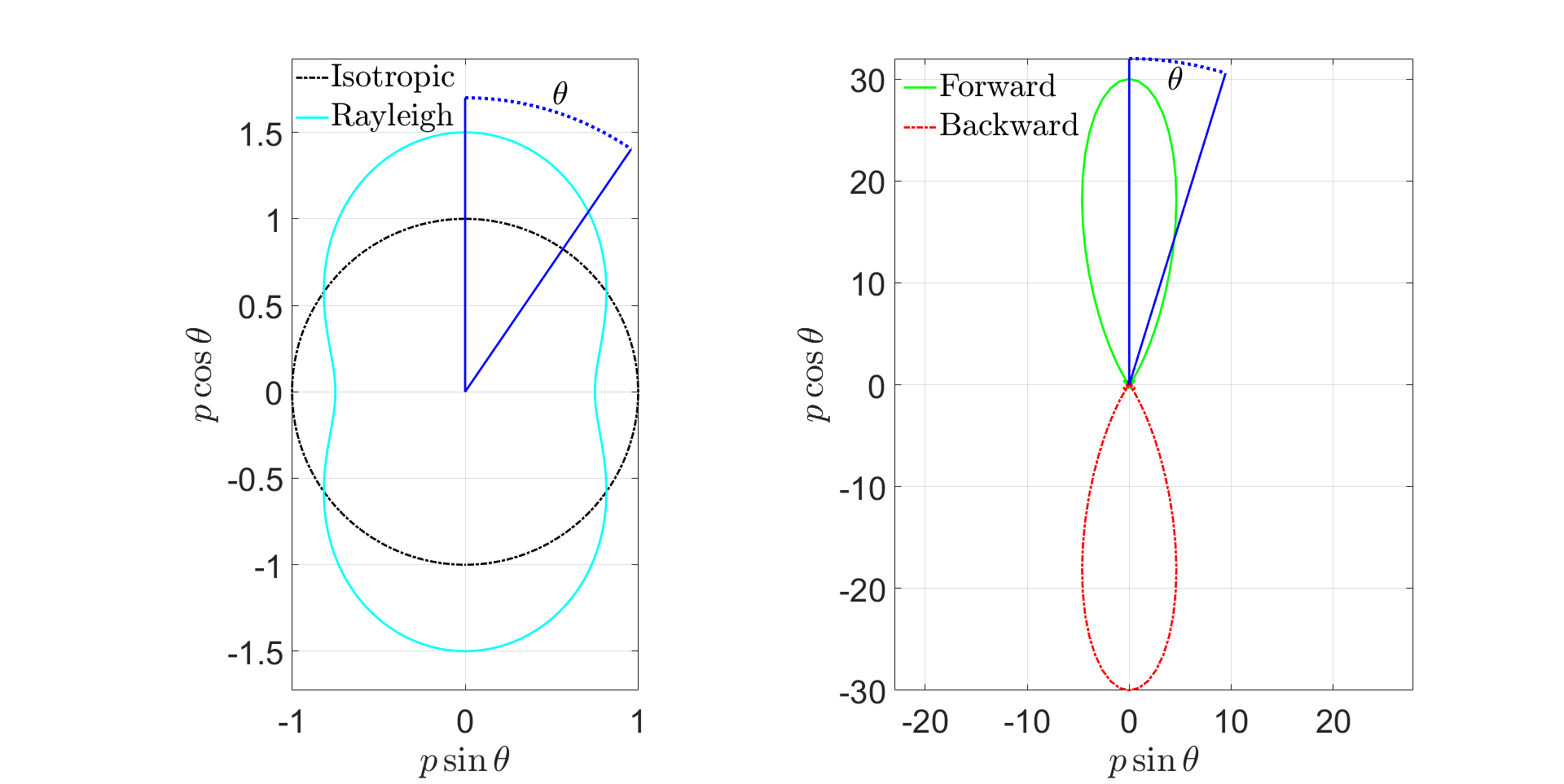}
\caption { The same scattering phase function
$p=p(\mu)$  of (\ref{spf10}) as for Fig. \ref{phase1} but plotted versus the direction cosine $\mu=\cos\theta$.
The forward-scattering and backward-scattering phase functions are described by polynomials in $\mu$ of degree $2m-1 = 9$, shown in the right panel.   The values of $p(\mu)$ for  forward  or backward scattering are $p(\pm 1)= m(m+1)=30$ and $p(\mp 1)=0$, respectively.}
\label{phase2}
\end{figure}

It is useful to display the phase function $p(\mu,\mu')$ for upward incoming radiation, with direction cosine $\mu'=1$. We can use (\ref{spf2}) to  write this reference phase function as
\begin{equation}
p(\mu)=p(\mu,1)=\sum_{l=0}^{2m-1} p_l(2l+1)P_l(\mu).
\label{spf10}
\end{equation}
Polar plots of the phase functions $p(\mu)=p(\cos\theta)$ of (\ref{spf10}) versus the zenith angle $\theta$ are shown in Fig. \ref{phase1} for the multipole coefficients $p_l$ of Table \ref{pl}. The same phase functions are plotted versus direction cosine $\mu =\cos\theta$ in Fig. \ref{phase2}.

\section{The  $2n$-Stream Equation of Transfer\label{int}}
The $2n$-stream equation of transfer (\ref{rtc6})
can be efficiently  solved with modern computer software. 
The angular dependence is characterized with $2n$ sample values of the intensity,
$I(\mu_i,\tau)\ge 0$, along the directions of the streams $i=1,2,3,\ldots, 2n$. As sketched in Fig. 1 of reference \cite{WH4},  the $i$th stream makes an angle $\theta_i=\cos^{-1}\mu_i$ to the zenith.  As shown by Fig. 2 of reference\,\cite{WH4},
the {\it Gauss-Legendre}\,\cite{Gauss} direction cosines, $\mu_i$,  are the zeros of the Legendre polynomial $P_{2n}$ of degree $2n$,  
\begin{equation}
P_{2n}(\mu_i)=0.
\label{int2}
\end{equation}
We will choose the indices $i=1,2,3,\cdots,2n$ such that
\begin{equation}
\mu_1<\mu_2<\mu_2<\cdots<\mu_{2n}.
\label{int4}
\end{equation}
Because the Legendre polynomial $P_{2n}$ is even, with $P_{2n}(\mu)=P_{2n}(-\mu)$,
the values of $\mu_i$ occur as equal and opposite pairs,
\begin{equation}
\mu_i=-\mu_{r(i)}.
\label{int6}
\end{equation}
The index reflection function is
\begin{equation}
r(i)=2n+1-i.
\label{int8}
\end{equation}
For the ordering convention (\ref{int4}) the indices $j$ for downward streams and the indices $k$ for upward streams are
\begin{eqnarray}
\mu_j&<0&\quad\hbox{for}\quad j=1,2,3,\ldots, n,\label{int9a}\\
\mu_k&>0&\quad\hbox{for}\quad k=n+1,n+2,n+3,\ldots, 2n.\label{int9b}
\end{eqnarray}
The weighted  sample values of the intensity, $w_i I(\mu_i,\tau)$, are denoted with the symbol
 $\lvec\mu_i|I(\tau)\}$,
\begin{equation}
\lvec\mu_i|I(\tau)\}=w_i I(\mu_i,\tau)\ge 0.
\label{int10}
\end{equation}
A formula for the Gauss-Legendre weights, $w_i>0$, was given by Eq. (13) of reference\,\cite{WH1} as
\begin{equation}
\frac{1}{w_i}=\frac{1}{w_{r(i)}}=\sum_{l=0}^{2n-1}\frac{2l+1}{2}P_l^2(\mu_i)>0.
\label{int12}
\end{equation}
We recall from Eq. (14) of reference\,\cite{WH1} that the weights sum to 2,
\begin{equation}
\sum_{i=1}^{2n}w_i=2.
\label{int13}
\end{equation}
We can think of the weights $w_i$ as discrete versions of direction-cosine increments, $w_i\approx \mu_i-\mu_{i-1}\to d\mu$ as $n\to \infty$.
\subsection{The $\mu$-space basis $|\mu_i)$\label{mu}}
To simplify subsequent equations, we write the weighted intensity samples, $w_i I(\mu_i)=\lvec\mu_i|I\}$,
as elements of an abstract vector $|I\}=|I(\tau)\}$. We can represent $|I\}$ with a $2n\times 1$ column  vector
\begin{equation}
|I\}=\sum_{i=1}^{2n}|\mu_i)\lvec \mu_i|I\}=\left[\begin{array}{c}\lvec\mu_1|I\}\\\lvec\mu_2|I\}\\ \vdots \\ \lvec\mu_{2n}|I\}\end{array}\right].
\label{int14}
\end{equation}
We will call the column vector on the right of  (\ref{int14}) the {\it $\mu$-space} representation of  the abstract vector $|I\}$.  We will use other $2n\times 1$ arrays of numbers to represent the same abstract vector $|I\}$ in other bases, for example, the multipole basis $|l)$, discussed in Section {\bf \ref{mm}}, or the $\lambda$-space bases $|\lambda_i)$ of Section {\bf \ref{ls}}.

The  {\it stream basis} vectors $|\mu_i)$ of (\ref{int14}) can be represented with the right (column) basis vectors
\begin{equation}
|\mu_1)=\left[\begin{array}{c}1\\0\\ \vdots \\ 0\end{array}\right],\quad|\mu_2)=\left[\begin{array}{c}0\\1\\ \vdots \\ 0\end{array}\right],\quad \cdots,\quad|\mu_{2n})=\left[\begin{array}{c}0\\0\\ \vdots \\ 1\end{array}\right].
\label{int16}
\end{equation}
Corresponding left (row) basis vectors can be represented as 
\begin{eqnarray}
\lvec\mu_1|&=&[1\quad 0\quad\cdots\quad 0],\nonumber\\
\lvec\mu_2|&=&[0\quad 1\quad\cdots\quad 0],\nonumber\\
&\vdots&\nonumber\\
\lvec\mu_{2n}|&=&[0\quad 0\quad\cdots\quad 1].
\label{int18}
\end{eqnarray}
We use a double left parenthesis, $\lvec\mu_i|$, as a reminder that  row basis vectors need not be Hermitian conjugates of column  basis vectors, although that happens to be the case for the stream basis vectors (\ref{int16}) and (\ref{int18}). One can think of column basis vectors as analogs of crystal lattice vectors and row vectors as analogs of reciprocal lattice vectors\,\cite{Reciprocal}.

As discussed in connection with Eq. (18) of reference\,\cite{WH4}, the stream basis vectors  are right and left eigenvectors of the direction cosine operator $\hat\mu$ 
\begin{equation}
\hat\mu|\mu_i)=\mu_i|\mu_i),\quad\hbox{and}\quad \lvec \mu_i|\hat\mu=\lvec \mu_i|\mu_i.
\label{int20}
\end{equation}
The eigenvectors $|\mu_i)$ and $\lvec\mu_i|$ have the orthonormality property
\begin{equation}
\lvec \mu_i|\mu_{i'})=\delta_{i i'}.
\label{int22}
\end{equation}
They have the completeness property 
\begin{equation}
\sum_{i=1}^{2n}|\mu_i)\lvec \mu_i|=\hat 1.
\label{int24}
\end{equation}
In (\ref{int24}) and elsewhere, we will use the symbol $\hat 1$ to denote a square identity operator with ones along the main diagonal and zeros elsewhere. $\hat 1$ has the same dimensions as any square operator that is added to, subtracted from,  or equated to it.
Multiplying (\ref{int24}) on the left or right by $\hat\mu$ and using (\ref{int20}) we find an expression for the direction-cosine operator,
\begin{equation}
\hat\mu =\sum_{i=1}^{2n}\mu_i|\mu_i)\lvec \mu_i|.
\label{int26}
\end{equation}
In (\ref{int26}) and elsewhere we will use the same symbol $\hat\mu$, to denote an operator and the matrix that represents the operator in some convenient basis.  For example, the matrix representing $\hat\mu$ in the $\mu$-space basis is diagonal and has the elements
\begin{equation}
\lvec\mu_i|\hat \mu|\mu_{i'})=\delta_{i i'}\mu_i.
\label{int27}
\end{equation}
 $2n$-stream computations must be done with  real numbers that are the elements of $2n\times 2n$ matrix representions of operators or the numerical elements of $2n\times 1$ or $1\times 2n$ column or row vectors.  But the basic theory can be described more clearly with operator equations like (\ref{rtc6}).

The direction secant operator $\hat\varsigma$ is the inverse of the direction cosine operator $\hat \mu$. We can use (\ref{int26}) to write
\begin{equation}
\hat\varsigma=\hat\mu^{-1} =\sum_{i=1}^{2n}\varsigma_i|\mu_i)\lvec \mu_i|.
\label{int28}
\end{equation}
The eigenvalues of the direction secant operator  $\hat \varsigma$ are the inverses of the eigenvalues 
$\mu_i$ of the direction cosine operator $\hat\mu$,
\begin{equation}
\varsigma_i=\frac{1}{\mu_i}.
\label{int30}
\end{equation}
The eigenvectors of (\ref{int20}) are also eigenvectors of $\hat\varsigma$,
\begin{equation}
\hat\varsigma|\mu_i)=\varsigma_i|\mu_i),\quad\hbox{and}\quad \lvec \mu_i|\hat\varsigma=\lvec \mu_i|\varsigma_i.
\label{int31}
\end{equation}

In accordance with Eq. (25) of reference\,\cite{WH4} it is convenient to use the stream basis vectors to define a projection operator $\mathcal{M}_{\bf d}$  for downward streams (\ref{int9a}) and a projection operator $\mathcal{M}_{\bf u}$ for upward streams (\ref{int9b}).
\begin{equation}
\mathcal{M}_{\bf d }=\sum_{j=1}^{n}|\mu_j)\lvec \mu_j|\quad\hbox{and}\quad
\mathcal{M}_{\bf u} =\sum_{k=n+1}^{2n}|\mu_k)\lvec \mu_k|.
\label{int38}
\end{equation}
In accordance with Eqs.  (29) of reference\, \cite{WH4}, we write the direction-cosine operator 
$\hat\mu$ as the sum of a downward part $\hat\mu_{\bf d}$ and an upward part $\hat\mu_{\bf u}$ in $\mu$-space,
\begin{equation}
\hat \mu =\hat\mu_{\bf d} +\hat\mu_{\bf u}.
\label{int52}
\end{equation}
According to Eq. (30) and Eq. (31) of reference\,\cite{WH4},
\begin{eqnarray}
\hat\mu_{\bf d} &=&\mathcal{M}_{\bf d}\hat\mu =\hat\mu\mathcal{M}_{\bf d} =\sum_{j=1}^n\mu_j|\mu_j)\lvec \mu_j|,\label{int54}\\
 \hat\mu_{\bf u} &=&\mathcal{M}_{\bf u}\hat\mu =\hat\mu\mathcal{M}_{\bf u} =\sum_{k=n+1}^{2n}\mu_k|\mu_k)\lvec \mu_k|.
\label{int56}
\end{eqnarray}

In like manner, we write the direction-secant operator 
$\hat\varsigma$ as the sum of a downward part $\hat\varsigma_{\bf d}$ and an upward part $\hat\varsigma_{\bf u}$ 
\begin{equation}
\hat \varsigma =\hat\varsigma_{\bf d} +\hat\varsigma_{\bf u}.
\label{int60}
\end{equation}
In analogy to (\ref{int56}),
expressions for the downward and upward parts are
\begin{eqnarray}
\hat\varsigma_{\bf d} &=&\mathcal{M}_{\bf d}\hat\varsigma =\hat\varsigma\mathcal{M}_{\bf d} =\sum_{j=1}^n\varsigma_j|\mu_j)\lvec \mu_j|,\label{int62}\\
 \hat\varsigma_{\bf u} &=&\mathcal{M}_{\bf u}\hat\varsigma =\hat\varsigma\mathcal{M}_{\bf u} =\sum_{k=n+1}^{2n}\varsigma_k|\mu_k)\lvec \mu_k|.
\label{int64}
\end{eqnarray}
The eigenvalues $\varsigma_i$ were given by (\ref{int30}).
\subsection{The multipole basis  $|l)$ \label{mm}}
Describing the angular distribution of the axially symmetric intensity $I(\mu, \tau)$ with the $2n$ sample
 values,  $I(\mu_i, \tau)$, at the  Gauss-Legendre direction cosines $\mu_i$ of  (\ref{int2}), is equivalent to approximating the intensity as a superposition of the first $2n$ Legendre polynomials,
\begin{equation}
I(\mu,\tau)= \sum_{l=0}^{2n-1}(2l+1)P_{l}(\mu)I_l(\tau).
\label {mm2}
\end{equation}
The $l$th intensity multipoles at the optical depth $\tau$  are
\begin{eqnarray}
I_l(\tau)&=&\lvec l|I(\tau)\}\nonumber\\
&=&\sum_{i=1}^{2n}\lvec l|\mu_i)\lvec\mu_i|I(\tau)\}.
\label {mm4}
\end{eqnarray}
Projections $\lvec l|\mu_i)$ of  the left multipole basis  $\lvec l|$ onto the right stream basis $|\mu_i)$, and vice versa, were given by Eqs.  (39) and (40) of reference\,\cite{WH4} in terms of the Legendre polynomials $P_l$, and the weights $w_i$ of (\ref{int12}) as
\begin{equation}
 \lvec l|\mu_i) =\frac{1}{2}P_l(\mu_i),
\label {mm6}
\end{equation}
and
\begin{equation}
\lvec\mu_i|l) = w_i(2l+1)P_{l}(\mu_i).
\label {mm8}
\end{equation}
Substituting (\ref{mm6}) and (\ref{int10}) into (\ref{mm4}), we find that the multipole moments $I_l(\tau)$ of the intensity are linear combinations of the sample values $I(\mu_i,\tau)$ at the 
 Gauss-Legendre direction cosines $\mu_i$,
\begin{equation}
I_l(\tau)
=\frac{1}{2}\sum_{i=1}^{2n}w_i P_l(\mu_i)I(\mu_i,\tau).
\label {mm9a}
\end{equation}
In analogy to (\ref{int22}) the multipole basis vectors $|l')$ and $\lvec l|$ have been chosen to have the orthonormality property
\begin{equation}
\lvec l|l')=\delta_{l l'}.
\label{mm9b}
\end{equation}
In analogy to (\ref{int24}), they  have the completeness property 
\begin{equation}
\sum_{l=0}^{2n-1}|l)\lvec l|=\hat 1.
\label{mm9c}
\end{equation}
Using (\ref{mm9b}) we can write the intensity vector as
\begin{equation}
|I(\tau)\}=\sum_{l=0}^{2n-1}|l)\lvec l|I(\tau)\}=\sum_{l=0}^{2n-1}|l)I_l(\tau).
\label{mm9d}
\end{equation}
The expansion coefficients $\lvec l|I(\tau)\}=I_l(\tau)$ were given by (\ref{mm9a}).

From (\ref{mm8}) we see that the elements $\lvec \mu_i|0)$ of the right monopole basis vector  $|0)$ are the weights $w_i$ of (\ref{int12})
\begin{equation}
|0)=\sum_{i=1}^{2n}|\mu_i)\lvec \mu_i|0)=\sum_{i=1}^{2n}|\mu_i)w_i=\left[\begin{array}{c}w_1\\ w_2\\ \vdots \\ w_{2n}\end{array}\right],\quad\hbox{or}\quad \lvec \mu_i|0)=w_i.
\label{mm10}
\end{equation}
The right monopole vector (\ref{mm10}) is particularly useful for representing the thermal emission source  vector $|B\}$ of (\ref{rtc6}),
\begin{equation}
|B\}=|0)B.
\label{mm11}
\end{equation}
Here $B$ is the Planck intensity of (\ref{rtc4}).

From (\ref{mm6}) we see that the elements $\lvec 0|\mu_i)$ of the left monopole basis vector $\lvec 0|$  are all equal to 1/2,
\begin{equation}
\lvec 0|=\sum_{i=1}^{2n}\lvec 0|\mu_i)\lvec \mu_i|=\frac{1}{2}\sum_{i=1}^{2n}\lvec \mu_i|=\bigg[\frac{1}{2}\quad \frac{1}{2}\quad\cdots \quad \frac{1}{2}\bigg]\quad\hbox{or}\quad \lvec 0|\mu_i)=\frac{1}{2}.
\label{mm12}
\end{equation}

To facilitate subsequent discussions, we note the identity from Eq. (48) of reference\,\cite{WH4}, 
\begin{equation}
\lvec 0|\hat \mu_{\bf u}^qe^{-\hat\varsigma_{\bf u}\tau}|0)=(-1)^q\lvec 0|\hat \mu_{\bf d}^qe^{\hat\varsigma_{\bf d}\tau}|0)=\frac{1}{2}E^{\{n\}}_{q+2}(\tau).
\label{mm20}
\end{equation}
Here $q$ is an integer,  $q=0$ or $q=1$ in our work. The $n$-stream exponential integral functions were given by Eq. (49) of reference\,\cite{WH4} as
\begin{equation}
E^{\{n\}}_{q}(\tau)=\sum_{k=n+1}^{2n}w_k\mu_k^{q-2}e^{-\tau/\mu_k},
\label{mm22}
\end{equation}
can be obtained by evaluating the exact exponential integral functions,
\begin{equation}
E_{q}(\tau)=\int_0^1d\mu\,  \mu^{q-2} e^{-\tau/\mu}, 
\label{mm24}
\end{equation}
with Gauss-Legendre quadratures\,\cite{Gauss}.   The exponential integral functions (\ref{mm24}) are discussed in Appendix I of Chandrasekhar's book\,\cite{Chandrasekhar}.  They account for the contributions to radiation transfer of intensity propagating at various slant angles with respect to the vertical. Graphical plots of the functions (\ref{mm24}) and its $n$-exponential approximation (\ref{mm22}) for $n\ge 5$ can hardly be distinguished, as shown by Fig. 9 of reference\,\cite{WH3}. 

For future reference we note from (\ref{mm20}) that 
\begin{equation}
\lvec 0|\hat \mu_{\bf u}|0)=\frac{1}{2}E^{\{n\}}_{3}(0)\to\frac{1}{4}\quad\hbox{as}\quad n\to \infty.
\label{mm30}
\end{equation}
The limit $E^{\{n\}}_{3}(0)\to 1/2$ as $n\to \infty$ was given by Eq. (222) of reference\,\cite{WH3}.
Some simple examples of (\ref{mm30}) are
\begin{equation}
\lvec 0|\hat \mu_{\bf u}|0)=\left \{\begin{array}{rl}0.2519, &\mbox{if $n=5$, }\\
0.2502,&\mbox{if $n=16$.} \end{array}\right . 
\label{mm32}
\end{equation}
\subsection{The $\lambda$-space basis $|\lambda_i)$ \label{ls}}
The exponentiation-rate operator $\hat \kappa$ of (\ref{rtc6}) was given by Eq.  (57) of reference\,\cite{WH4} as
\begin{equation}
\hat \kappa = \hat\varsigma\hat\eta,
\label{ls2}
\end{equation}
the product of the direction-secant operator $\hat\varsigma$ of  (\ref{int28}), and 
the efficiency operator 
\begin{equation}
\hat \eta =\hat 1-\frac{1}{2}\tilde\omega \hat p,
\label{ls4}
\end{equation}
given by Eq. (58) of reference\,\cite{WH4}.
In accordance with Eq. (61) of reference\,\cite{WH4},
the scattering operator $\hat p$ of (\ref{ls4}) can be written in terms of  the right and left multipole basis vectors, $|l)$ and $\lvec l|$  of (\ref{mm6}) and (\ref{mm8}) and the multipole coefficients $p_l$ of (\ref{spf2}) as
\begin{equation}
\hat p = 2\sum_{l=0}^{2n-1}p_l|l)\lvec l|.
\label{ls6}
\end{equation}
 For a non-scattering atmosphere, when the single-scattering albedo vanishes, $\tilde\omega \to 0$, the efficiency operator $\hat\eta$ of (\ref{ls4}) reduces to the identity operator, $\hat \eta\to \hat 1$ and the exponentiation-rate operator $\hat\kappa$ of (\ref{ls2}) reduces to the direction secant operator $\hat\kappa \to \hat \varsigma$ of (\ref{int28}). So $\hat\kappa$ is a generalized direction secant operator $\hat\varsigma$ for a scattering atmosphere.  

It is convenient to introduce a penetration-length operator, 
\begin{equation}
\hat \lambda =\hat\kappa^{-1} =\hat\eta^{-1}\hat\mu,
\label{ls8}
\end{equation}
the inverse of the exponentiation-rate operator $\hat \kappa$ of (\ref{ls2}). From (\ref{ls8}) and (\ref{ls4}) we see that $\hat \lambda\to\hat\mu$ as $\tilde\omega \to 0$ and $\hat\eta^{-1}\to \hat 1$.   So $\hat\lambda$ is a generalized direction cosine operator $\hat \mu$ for a scattering atmosphere. 

We denote right and left eigenvectors, $|\lambda_i)$ and $\lvec\lambda_i|$ of the penetration-length operator $\hat\lambda$, in analogy to (\ref{int20}), by
\begin{equation}
\hat \lambda |\lambda_i)=\lambda_i |\lambda_i)\quad\hbox{and}\quad\lvec\lambda_i|\hat \lambda = \lvec\lambda_i|\lambda_i.
\label{ls10}
\end{equation}
As in (\ref{int4})
the real eigenvalues or {\it penetration lengths} are ordered such that 
\begin{equation}
\lambda_1<\lambda_2<\lambda_3<\cdots\lambda_{2n}.
\label{sm4}
\end{equation}
As in (\ref{int6}) the penetration lengths for a homogeneous cloud have the reflection symmetry
\begin{equation}
\lambda_i=-\lambda_{r(i)}.
\label{sm6}
\end{equation}
The index reflection function $r(i)=2n+1-i$ was defined by (\ref{int8}). The {\it $\lambda$-space} basis vectors $\lvec \lambda_i|$ and $|\lambda_j)$ are chosen to have orthonormality and completeness relations analogous to (\ref{int22}) and (\ref{int24}),
\begin{equation}
\lvec \lambda_i|\lambda_j)=\delta_{ij},
\label{sm8}
\end{equation}
and 
\begin{equation}
\hat 1 =\sum_{i=1}^{2n}|\lambda_i)\lvec \lambda_i|.
\label{sm9a}
\end{equation}
In analogy to (\ref{int38}) we define the downward and upward projection operators in $\lambda$ space by
\begin{equation}
\mathcal{L}_{\bf d}=\sum_{j=1}^{n}|\lambda_j)\lvec \lambda_j|\quad\hbox{and}\quad
\mathcal{L}_{\bf u} =\sum_{k=n+1}^{2n}|\lambda_k)\lvec \lambda_k|.
\label{sm10}
\end{equation}
The exponentiation-rate operator (\ref{ls2}) can be written as the sum of downward and upward parts in $\lambda$ space
\begin{equation}
\hat\kappa=\hat\kappa_{\bf d}+\hat\kappa_{\bf u}. 
\label{sm12}
\end{equation}
where
\begin{equation}
\hat\kappa_{\bf d}=\sum_{j=1}^{n}\kappa_j|\lambda_j)\lvec \lambda_j|\quad\hbox{and}\quad
\hat\kappa_{\bf u} =\sum_{k=n+1}^{2n}\kappa_k|\lambda_k)\lvec \lambda_k|.
\label{sm14}
\end{equation}
The eigenvalues $\kappa_i$ of $\hat\kappa$ are the inverses of the eigenvalues $\lambda_i$ of $\hat\lambda$, and the eigenvectors are the same as those of $\hat\lambda$,
\begin{equation}
\kappa_i=\frac{1}{\lambda_i},\quad \lvec\kappa_i|= \lvec\lambda_i|,\quad\hbox{and}\quad |\kappa_i)=|\lambda_i).
\label{sm16}
\end{equation}
The $\lambda$-space basis is especially useful for homogeneous clouds, where the exponentiation-rate operator $\hat\kappa$ of (\ref{ls2}) is independent of the optical depth $\tau$.
\section{Clouds\label{cl}}
In this section we show how to use $2n$-stream radiative theory\,\cite{WH4} to model the thermal emission of clouds that have greenhouse gases in the air between the condensed-phase particulates, water droplets or ice crystallites.
\subsection{The scattering operator $\mathcal{S}$ \label{sm}}
Eq. (104) of reference\,\cite{WH4} shows that a cloud scatters external incoming intensity vector $|\ddot I^{\{\rm in}\}$  into outgoing intensity vector $|\ddot I^{\{\rm out}\}$ as described by the scattering  operator $\mathcal{S}$,
\begin{equation}
|\ddot I^{\{\rm out}\}=\mathcal{S}|\ddot I^{\{\rm in}\}.
\label{s2}
\end{equation}
Eq. (109) of reference\,\cite{WH4} shows that
the emissivity operator $\mathcal{E}$ describes
the outgoing intensity $|\dot I^{\{\rm out}\}$ produced by thermal emission of the cloud, when it has a uniform temperature $T$ and the corresponding Planck intensity $B$ of (\ref{rtc4}),
\begin{equation}
|\dot I^{\{\rm out}\}=\mathcal{E}|0)B.
\label{s4}
\end{equation}
The monopole basis $|0)$ was given by (\ref{mm10}). As discussed in Section {\bf 2.6} of reference\,\cite{WH4}, the double dots of $|\ddot I^{\{\rm out}\}$ and $|\ddot I^{\{\rm in}\}$ in (\ref{s2}) identify intensities orginating from sources outside the cloud. The single dot of $|\dot I^{\{\rm out}\}$ in (\ref{s4}) identifies intensities generated by thermal emission of cloud particulates and greenhouse gases.

According to Kirchhoff's law, Eq. (111) of reference \,\cite{WH4}, the emissivity operator $\mathcal{E}$ of (\ref{s4}) and the scattering operator  $\mathcal{S}$ of (\ref{s2}) sum to the $2n\times 2n$ identity operator $\hat 1$,
\begin{equation}
\mathcal{E} +\mathcal{S}=\hat 1.
\label{s6}
\end{equation}

\subsection{Intensity emissivities\label{ie}}

Using (\ref{int10}), (\ref{mm10}) and (\ref{s4}), we write the outgoing thermal intensity of the $i$th stream from an isothermal cloud as
\begin{equation}
\dot I^{(\rm out)}_i=\frac{\lvec\mu_i|\dot I^{(\rm out)}\}}{w_i}=\frac{\lvec\mu_i|\mathcal{E}|0)B}{\lvec\mu_i|0)}.
\label{ie2}
\end{equation}
 We can write (\ref{ie2}) as
\begin{equation}
\dot I^{(\rm out)}_i=\varepsilon_i B, 
\label{ie4}
\end{equation}
where the intensity emissivity $\varepsilon_i$ is 
\begin{equation}
\varepsilon_i=\frac{\lvec\mu_i|\mathcal{E}|0)}{\lvec\mu_i|0)}=1-\frac{\lvec\mu_i|\mathcal{S}|0)}{\lvec\mu_i|0)}.
\label{ie6}
\end{equation}
 \subsection{Flux emissivities\label{fe}}
According to (\ref{s4}) and Eq. (90) of reference\,\cite{WH4}, the upward flux vector
$|\dot Z^{(\rm out)}_{\bf u}\}$ for an isothermal cloud is
\begin{equation}
|\dot Z^{(\rm out)}_{\bf u}\}=4\pi \hat\mu_{\bf u}|\dot I^{(\rm out)}\}=4\pi\hat\mu_{\bf u}\mathcal{E}|0)B,\label{fe2}
\end{equation}
According to Eq. (69) of reference\,\cite{WH4}, the scalar flux corresponding to (\ref{fe2}) is
\begin{equation}
\dot Z^{(\rm out)}_{\bf u}=\lvec 0|\dot Z^{(\rm out)}_{\bf u}\}=4\pi\lvec 0|\hat\mu_{\bf u}\mathcal{E}|0)B.\label{fe4}
\end{equation}
For a black cloud, where the emissivity operator is $\mathcal{E}=\hat 1$, the outgoing upward flux follows from (\ref{fe4}) and is
\begin{eqnarray}
\dot Z^{(\rm bb)}&=&4\pi\lvec 0|\hat\mu_{\bf u}|0)B\nonumber\\
&\to&\pi B\quad\hbox{as}\quad n\to\infty.
\label{fe6}
\end{eqnarray}
We used (\ref{mm30}) to write the last line of (\ref{fe6}).
From inspection of (\ref{fe4}) and (\ref{fe6}) we see that we can write
\begin{equation}
\dot Z^{(\rm out)}_{\bf u}=\varepsilon_{\bf u}\dot Z^{(\rm bb)}, \label{fe8}
\end{equation}
where we use (\ref{fe4}), (\ref{fe6})  and (\ref{s6}) to write the upward {\it flux emissivity} of (\ref{fe8}) as 
\begin{equation}
\varepsilon_{\bf u} =\frac{\dot Z^{(\rm out)}_{\bf u}}{\dot Z^{(\rm bb)}}=\frac{\lvec 0|\hat\mu_{\bf u}\mathcal{E}|0)}{ \lvec 0|\hat\mu_{\bf u}|0)}=1-\frac{\lvec 0|\hat\mu_{\bf u}\mathcal{S}|0)}{ \lvec 0|\hat\mu_{\bf u}|0)}.
\label{fe10}
\end{equation} 
In analogy to (\ref{fe8}) we can write the downward flux from an isothermal cloud as
\begin{equation}
\dot Z^{(\rm out)}_{\bf d}=\varepsilon_{\bf d}\dot Z^{(\rm bb)}. \label{fe14}
\end{equation}
The downward flux emissivity is 
\begin{equation}
\varepsilon_{\bf d} =\frac{\dot Z^{(\rm out)}_{\bf d}}{\dot Z^{(\rm bb)}}=\frac{\lvec 0|\hat\mu_{\bf d}\mathcal{E}|0)}{ \lvec 0|\hat\mu_{\bf d}|0)}=1-\frac{\lvec 0|\hat\mu_{\bf d}\mathcal{S}|0)}{ \lvec 0|\hat\mu_{\bf d}|0)}.
\label{fe16}
\end{equation}
It is straightforward to show that the upward and downward flux emissivities $\varepsilon_{\bf u}$  and $\varepsilon_{\bf d}$ of (\ref{fe10})  and (\ref{fe16}) are related to the intensity emissivities $\varepsilon_i$ of (\ref{ie4}) by
\begin{equation}
\varepsilon_{\bf d} =\frac{\sum_j \mu_j w_j \varepsilon_j}{\sum_j\mu_j w_j}
\quad\hbox{and}\quad\varepsilon_{\bf u} =\frac{\sum_k \mu_k w_k \varepsilon_k}{\sum_k\mu_k w_k}.
\label{fe18}
\end{equation}
The summation indices $j$ and $k$ of (\ref{fe18}) for downward and upward streams were given by (\ref{int9a}) and (\ref{int9b}).

In Eq. (163)  of reference\,\cite{WH4} we showed that the downward and upward flux albedos $\omega_{\bf d}$  and $\omega_{\bf u}$ of an isothermal cloud are the complements of the downward and upward flux  emissivities $\varepsilon_{\bf d}$ and  $\varepsilon_{\bf u}$ of (\ref{fe16}) and  (\ref{fe10}),
\begin{equation}
\omega_{\bf d}=1-\varepsilon_{\bf d}\quad\hbox{and}\quad\omega_{\bf u}=1-\varepsilon_{\bf u}.
\label{fe24}
\end{equation}
Anything that increases the downward or upward flux emissivities,    $\varepsilon_{\bf d}$ and $\varepsilon_{\bf u}$, will decrease the corresponding downward and upward flux albedos, $\omega_{\bf d}$ and $\omega_{\bf u}$, and vice versa.
\subsection{Bounds on emissivities \label{be}}
In Eq. (117) of reference\,\cite{WH4} we mentioned the fundamental bounds on the scattering operator $\mathcal{S}$
\begin{equation}
0\le \lvec \mu_i|\mathcal{S}|0)\le \lvec \mu_i|0).
\label{be2}
\end{equation}
If we replace  $\mathcal{S}$ in (\ref{be2}) by $\hat 1-\mathcal{E}$ from  Kirchhoff's law (\ref{s6}) we find
that the emissivity operator $\mathcal{E}$ satisfies an analogous bounding equation
\begin{equation}
0\le \lvec \mu_i|\mathcal{E}|0)\le \lvec \mu_i|0).
\label{be4}
\end{equation}
Dividing all terms of (\ref{be4}) by the factor $\lvec \mu_i|0)=w_i>0$ from (\ref{mm10}) we find that the intensity emissivity  $\varepsilon_i$ of (\ref{ie6}) is bounded by
\begin{equation}
0\le \varepsilon_i\le 1.
\label{be6}
\end{equation}
Using the bounds (\ref{be6}) in (\ref{fe18}) we see that the values of the upward and downward flux emissivities are also constrained to the same interval,
\begin{equation}
0\le \varepsilon_{\bf d}\le 1\quad\hbox{and}\quad 0\le \varepsilon_{\bf u}\le 1.
\label{be8}
\end{equation}

\subsection{Homogeneous, optically-thick clouds\label{hc}}
To simplify further discussion of the intensity emissivities $\varepsilon_i$ of (\ref{ie6}) and flux emissivities, $\varepsilon_{\bf u}$ and $\varepsilon_{\bf d}$ of (\ref{fe10}) and (\ref{fe16}),  we will consider optically-thick homogeneous clouds, that is, clouds with the same single-scattering albedo $\tilde\omega$ and the same scattering phase function (\ref{spf2}) at every location inside.
According to Eq. (232) of reference\,\cite{WH1},  we can write the scattering operator  $\mathcal{S}$ for an optically-thick, homogeneous cloud in the limit $\tau\to\infty$ as
\begin{equation}
\mathcal{S}=\mathcal{C}_{\bf d u}(\mathcal{C}_{\bf u u})^{-1}+\mathcal{C}_{\bf u d}(\mathcal{C}_{\bf d d})^{-1}.
\label{hm2}
\end{equation}
Here and elsewhere we will use the same symbol, a negative one superscript ($^{-1}$), to denote pseudo-inverse operators\,\cite{pseudoinverse},  like $\mathcal{C}_{\bf d d}^{-1}$ or $\mathcal{C}_{\bf u u}^{-1}$ of (\ref{hm2}), for which  $\mathcal{C}_{\bf d d}^{-1}\mathcal{C}_{\bf d d}= \mathcal{M}_{\bf d }$ or
 $\mathcal{C}_{\bf u u}^{-1}\mathcal{C}_{\bf u u}= \mathcal{M}_{\bf u }$,
and true inverse operators like $\hat\mu^{-1}=\hat\varsigma$ of (\ref{int28}), for which 
$\hat\mu^{-1}\hat\mu=\hat 1 =\mathcal{M}_{\bf u }+\mathcal{M}_{\bf d }$.
 The context will make the distinction clear in cases where it matters.

In (\ref{hm2})
the {\it overlap} operators $\mathcal{C}_{\bf q q'}$ between $\mu$-space projection operators $\mathcal{M}_{\bf q}$ of (\ref{int38}) and the $\lambda$-space projection operators $\mathcal{L}_{\bf q'}$ of (\ref{sm10}), are defined by Eq. (131) of reference\, \cite{WH4} as
\begin{equation}
\left[\begin{array}{ll}\mathcal{C}_{\bf d d}&\mathcal{C}_{\bf d u}\\ \mathcal{C}_{\bf u d}&\mathcal{C}_{\bf u u}\end{array}\right]=\left[\begin{array}{ll}\mathcal{M}_{\bf d}\mathcal{L}_{\bf d}&\mathcal{M}_{\bf d}\mathcal{L}_{\bf u}\\ \mathcal{M}_{\bf u}\mathcal{L}_{\bf d}&\mathcal{M}_{\bf u}\mathcal{L}_{\bf u}\end{array}\right].
\label{hm4}
\end{equation}
The overlap operators  $\mathcal{C}_{\bf q q'}$ of (\ref{hm4}) depend on the single-scattering albedo $\tilde\omega$ and on the multipole coefficients $p_l$ of (\ref{spf2}) for the scattering phase function. 

\begin{figure}[t]
\includegraphics[height=100mm,width=1\columnwidth]{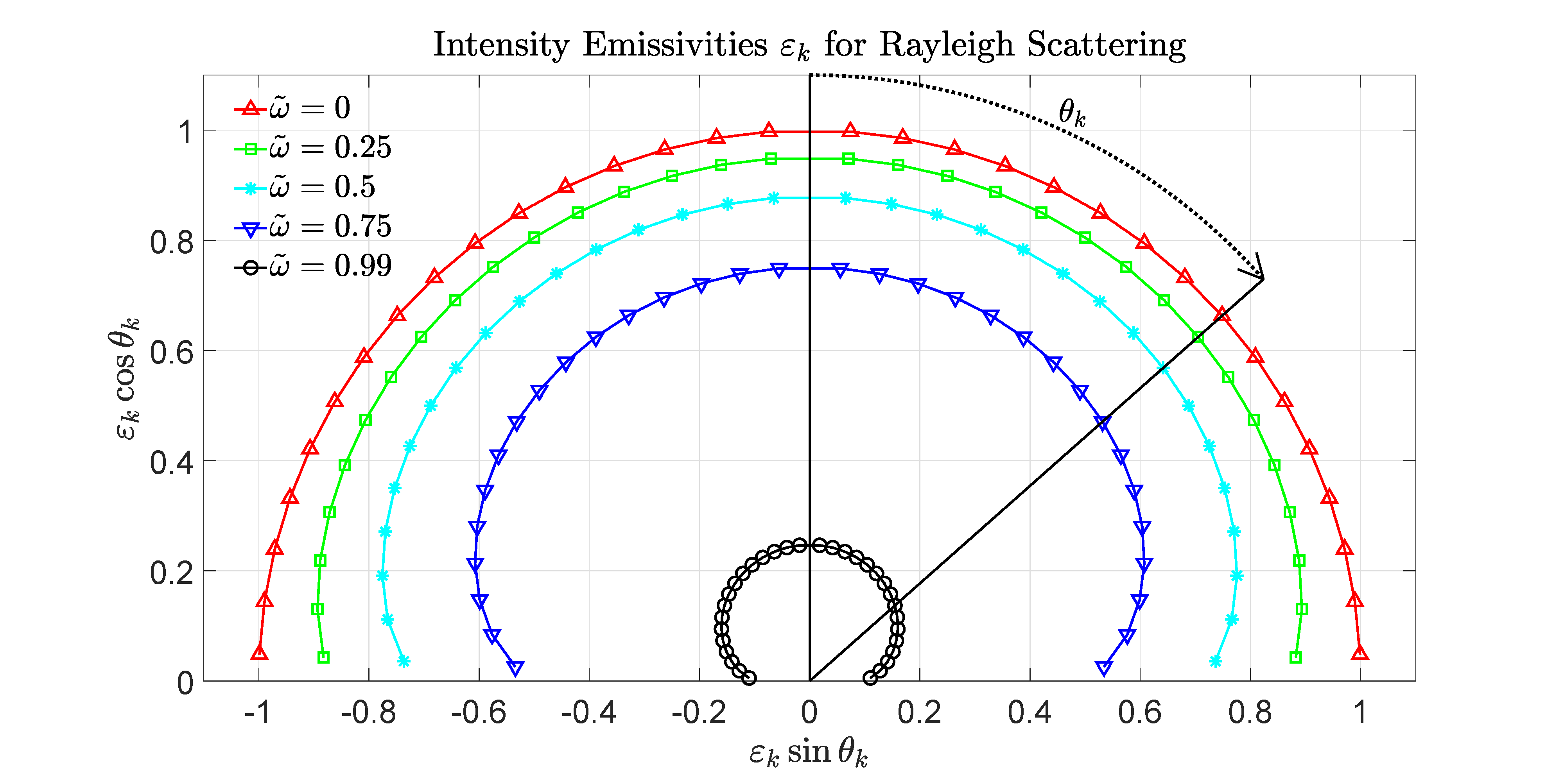}
\caption { The intensity emissivities $\varepsilon_k$ of (\ref{hm6}) for an optically thick cloud with Raleigh scattering, modeled with $2n = 32$ streams. The scattering phase function $p(\mu)$ is shown in the left panel of Fig. \ref{phase1}. The multipole coefficients $p_l$ of the phase function are given in Table \ref{pl}.}
\label{aemis2}
\end{figure}

\begin{figure}[t]
\includegraphics[height=100mm,width=1\columnwidth]{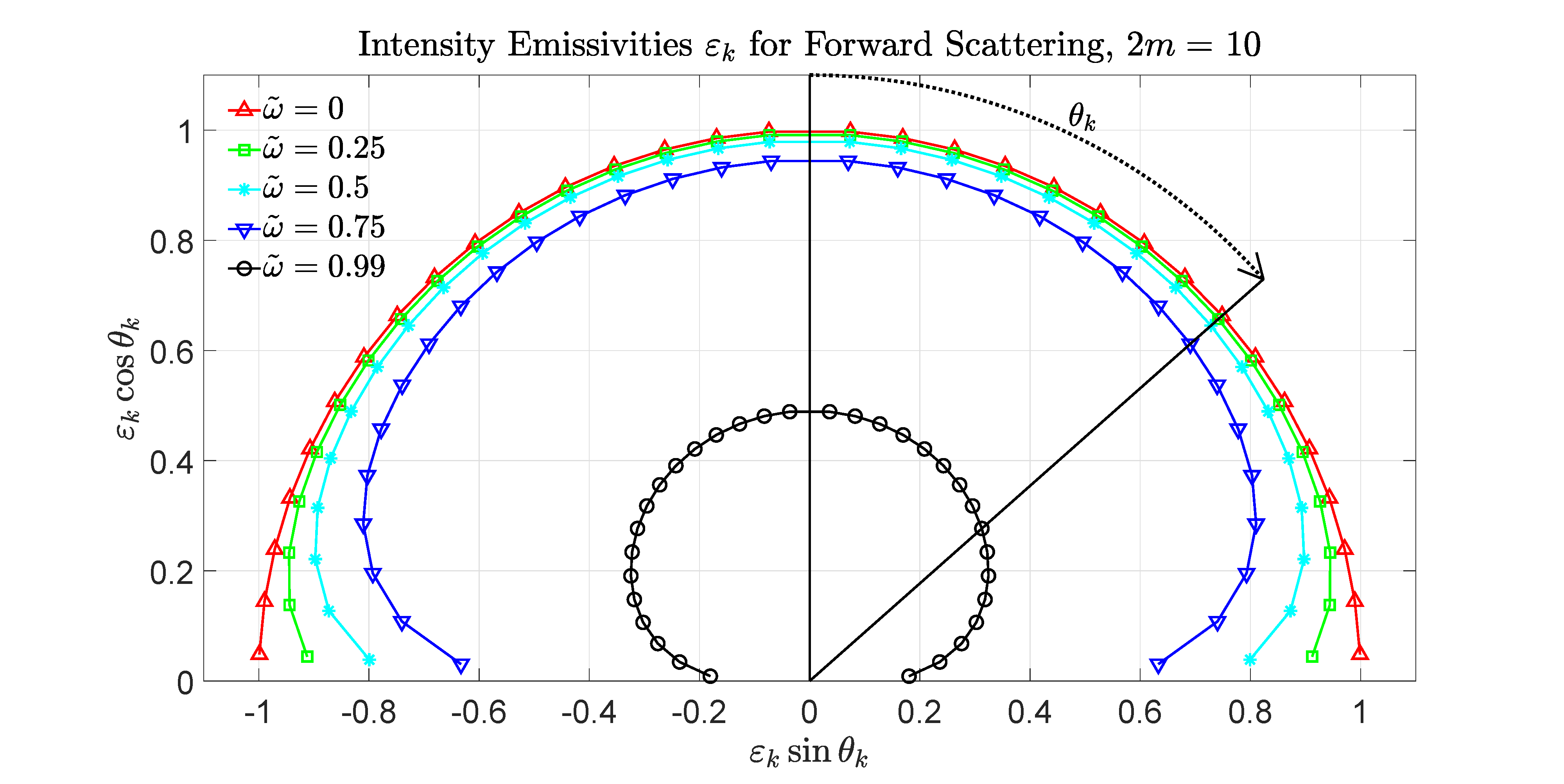}
\caption {Like Fig. \ref{aemis2} but for the maximum forward-scattering phase function $p(\mu)$ of (\ref{spf10}) that can be constructed from the first $2m=10$ Legendre polynomials. The scattering phase function $p(\mu)$ is shown in the right panel of Fig. \ref{phase1} or Fig. \ref{phase2}.}
\label{aemis3}
\end{figure}

\begin{figure}[t]
\includegraphics[height=100mm,width=1\columnwidth]{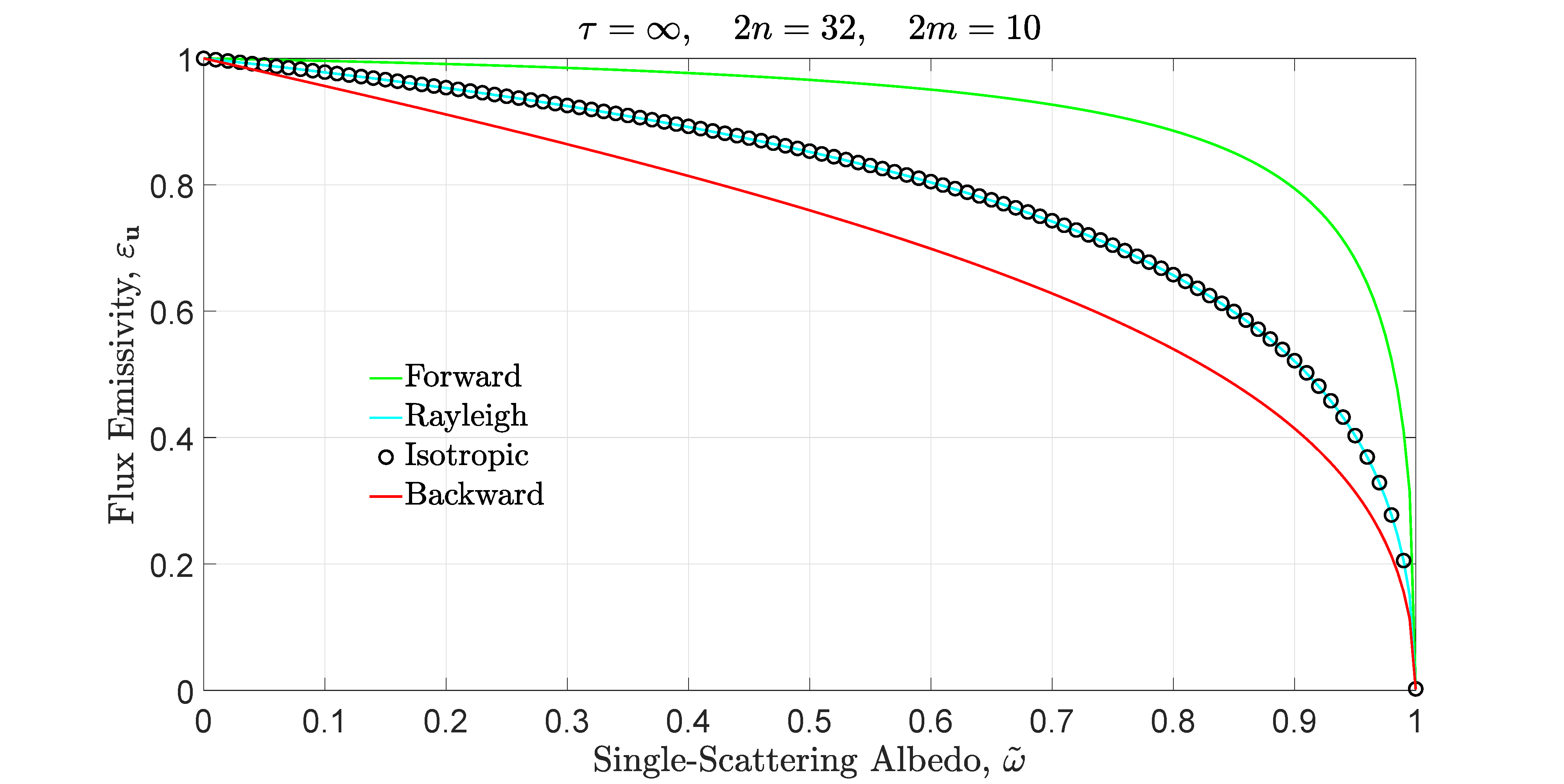}
\caption {Upward-flux emissivities $\varepsilon_{\bf u}$ of (\ref{fe10}) versus the single-scattering albedo $\tilde \omega$ for homogeneous, optically thick clouds with the scattering-phase operators  $\hat p$ of (\ref{ls6}),  evaluated with the multipole coefficients $p_l$ of Table \ref{pl}.}
\label{emis2}
\end{figure}

Substituting (\ref{hm2}) into the formula (\ref{ie6}) and noting that $\lvec \mu_{k}|\mathcal{C}_{\bf d u}=\breve 0$, we find that the upward stream emissivities are
\begin{equation}
\varepsilon _{k}=
1-\frac{\lvec \mu_{k}|\mathcal{C}_{\bf u d}(\mathcal{C}_{\bf d d})^{-1}|0)}{ \lvec \mu_{k}|0)}.
\label{hm6}
\end{equation}
The indices $k$ for upward streams were given by (\ref{int9b}).
Substituting (\ref{hm2}) into (\ref{fe10}), and noting that $\hat\mu_{\bf u}\mathcal{C}_{\bf d u}=\breve 0$, we find that the upward flux emissivity is
\begin{equation}
\varepsilon _{\bf u}=
1-\frac{\lvec 0|\hat\mu_{\bf u}\mathcal{C}_{\bf u d}(\mathcal{C}_{\bf d d})^{-1}|0)}{ \lvec 0|\hat\mu_{\bf u}|0)}.
\label{hm8}
\end{equation}
For the homogeneous clouds considered in this section there is up-down symmetry, so that
\begin{equation}
\varepsilon _{j}=\varepsilon _{r(k)}=\varepsilon_{k}\quad \hbox{and} \quad\varepsilon _{\bf d}=\varepsilon _{\bf u}.
\label{hm10}
\end{equation}
In (\ref{hm10}) the index, $j=r(k) =2n+1-k \le n$ of (\ref{int8}), labels the downward stream, with direction cosine $\mu_j=-\mu_k$,  the negative of the positive  direction cosine $\mu_k>0$ of the upward stream with the index $k>n$.

Emissivities like those of (\ref{ie6}), (\ref{fe10}) or (\ref{fe16})  for general, inhomogeneous clouds, need not have up-down symmetry. For example,  as one can see from the discussions of cloud stacks in reference \,\cite{WH4}, or from intuitive physical considerations,  a compound cloud consisting of a scattering  cloud directly above a black cloud must have $\epsilon_{\bf u} <1$ and $\epsilon_{\bf d} = 1$.

Fig. \ref{aemis2} shows representative intensity emissivities $\varepsilon_k$ of  (\ref{hm6}) for a homogeneous, optically thick cloud where the particulates have  Rayleigh scattering phase functions, like those in the left panels of Fig. \ref{phase1} and Fig. \ref{phase2}. The multipole coefficients $p_l$ of the phase function $\hat p$ are given in Table \ref{pl}.  The cloud is modeled with $2n = 32$ streams. The red curve with $\tilde\omega=0$ shows the Lambertian (uniform) angular distribution of  stream emissivities $\varepsilon_k$ of a black body.  As the single-scattering albedo $\tilde \omega$ increases the intensity emissivities $\varepsilon_k$ decrease and become increasingly limb darkened.

The reason for limb darkening of isothermal clouds like those of Fig. \ref{aemis2} is not the same as for limb darkening of the Sun. In the Sun's photosphere\,\cite{photosphere}, the single-scattering albedo is nearly zero,  $\tilde \omega\approx 0$, since the negative hydrogen ions, H$^{-}$, which are responsible for most of the visible opacity, have relatively large absorption cross sections  and relatively small scattering cross sections, much like greenhouse gases for thermal radiation in Earth's atmosphere.  If the Sun were isothermal from the photosphere to several e-foldings of optical depth (a few hundred km) inward, the angular distribution of emission would be nearly Lambertian, like the red curve of Fig. \ref{aemis2},  and there would be little limb darkening.

Fig. \ref{aemis3} is analogous to Fig. \ref{aemis2} except it is calculated for a forward scattering phase function $p(\mu)$, like those in the right panels of Fig. \ref{phase1} and Fig. \ref{phase2}. Other things being equal, forward scattering clouds remain more nearly black, with high emissivities, for much larger values of the single-scattering albedo $\tilde\omega$, than is the case for Rayleigh scattering, which is nearly isotropic.  Forward scattered photons can penetrate much deeper into the optically thick cloud and have a larger probability of being absorbed before being multiply scattered back out of the input surface.

Fig. \ref{emis2} shows  how the single scattering albedo, $\tilde \omega$, modifies the upward flux emissivity $\varepsilon_{\bf u}$ of (\ref{hm8}) for fixed values of the scattering-phase operator $\hat p$ of (\ref{ls6}), evaluated with the multipole coefficients $p_l$ of Table \ref{pl}. For clouds of water droplets or ice crystallites, single-scattering albedos $\tilde \omega > 0.9$ are observed  for visible light. For example, see 
Fig.\,{\bf 12.10} on page 374 of reference\,\cite{Petty}. Smaller values of the single-scattering albedo, say $\tilde \omega \sim 1/2$, appear to characterize the scattering of thermal radiation by cloud particulates. But detailed observational data on scattering and absorption of thermal radiation by cloud particulates is very sparse compared to  that for visible light.

\subsection{Frequency-integrated emissivities}
The frequency-integrated  flux (\ref{fe4}) out of the top of an isothermal cloud is
\begin{eqnarray}
\langle \dot Z_{\bf u}^{(\rm out)}\rangle&=&\int_0^{\infty}d\nu\,\dot Z^{(\rm out)}_{\bf u}\nonumber\\
&=&4\pi\int_0^{\infty}d\nu\,\lvec 0|\hat\mu_{\bf u}\mathcal{E}|0)B.
\label{fie2}
\end{eqnarray}
Here the Planck intensity $B$ of (\ref{rtc4}) is a function of frequency $\nu$, and we assume that the emissivity operator $\mathcal{E}$ may also be a function of frequency. A black cloud has a frequency-independent emissivity operator $\mathcal{E} = \hat 1$, so we can use (\ref{fie2}) to write the frequency-integrated flux emerging from the top or bottom of an isothermal black cloud as
\begin{eqnarray}
\langle\dot Z^{(\rm bb)}\rangle
&=&4\pi \lvec 0|\hat\mu_{\bf u}|0)\int_0^{\infty}d\nu\,B\nonumber\\
&=&4 \lvec 0|\hat\mu_{\bf u}|0)\sigma_{\rm SB}T^4\nonumber\\
&\to&\sigma_{\rm SB}T^4\quad\hbox{as}\quad n\to\infty.
\label{fie4}
\end{eqnarray}
The second line of (\ref{fie4}) comes from the fact that the frequency integral of the Planck intensity (\ref{rtc4}) is
\begin{equation}
\int_0^{\infty}d\nu B=\frac{\sigma_{\rm SB}T^4}{\pi}.
\label{fie6}
\end{equation}
Here $T$ is the absolute cloud temperature, and the Stefan-Boltzmann constant is
\begin{equation}
\sigma_{\rm SB}=\frac{2\pi^5k_{\rm B}^4}{15c^2h_{\rm P}^3}=5.669 \times 10^{-5} \hbox{ erg s$^{-1}$ cm$^{-2}$ K$^{-4}$}.
\label{fie8}
\end{equation}
The last line of (\ref{fie4}) follows from (\ref{mm30}).
Comparing (\ref{fie2}) with (\ref{fie4}) we write
\begin{eqnarray}
\langle \dot Z_{\bf u}^{(\rm out)}\rangle=\bar\varepsilon_{\bf u}\langle\dot Z^{(\rm bb)}\rangle. 
\label{fie10}
\end{eqnarray}
Solving (\ref{fie10}) for
the frequency averaged emissivity $\bar\varepsilon_{\bf u}$ and using (\ref{fie2}) and (\ref{fie4}) we find
\begin{eqnarray}
\bar\varepsilon_{\bf u}&=&\frac{\langle\dot Z_{\bf u}^{(\rm out)}\rangle}{\langle \dot Z^{(\rm bb)}\rangle}\nonumber\\
&=&
\frac{\pi}{\lvec 0|\hat\mu_{\bf u}|0)\sigma_{\rm SB}T^4}\int_0^{\infty}d\nu \lvec 0|\hat\mu_{\bf u}\mathcal{E}|0)B\nonumber\\
&=&\frac{\pi}{\sigma_{\rm SB}T^4}\int_0^{\infty}d\nu\, \varepsilon_{\bf u}B.
\label{fie12}
\end{eqnarray}

\subsection{Cloud emissivities at different altitudes}

\begin{figure}[t]
\includegraphics[height=100mm,width=1\columnwidth]{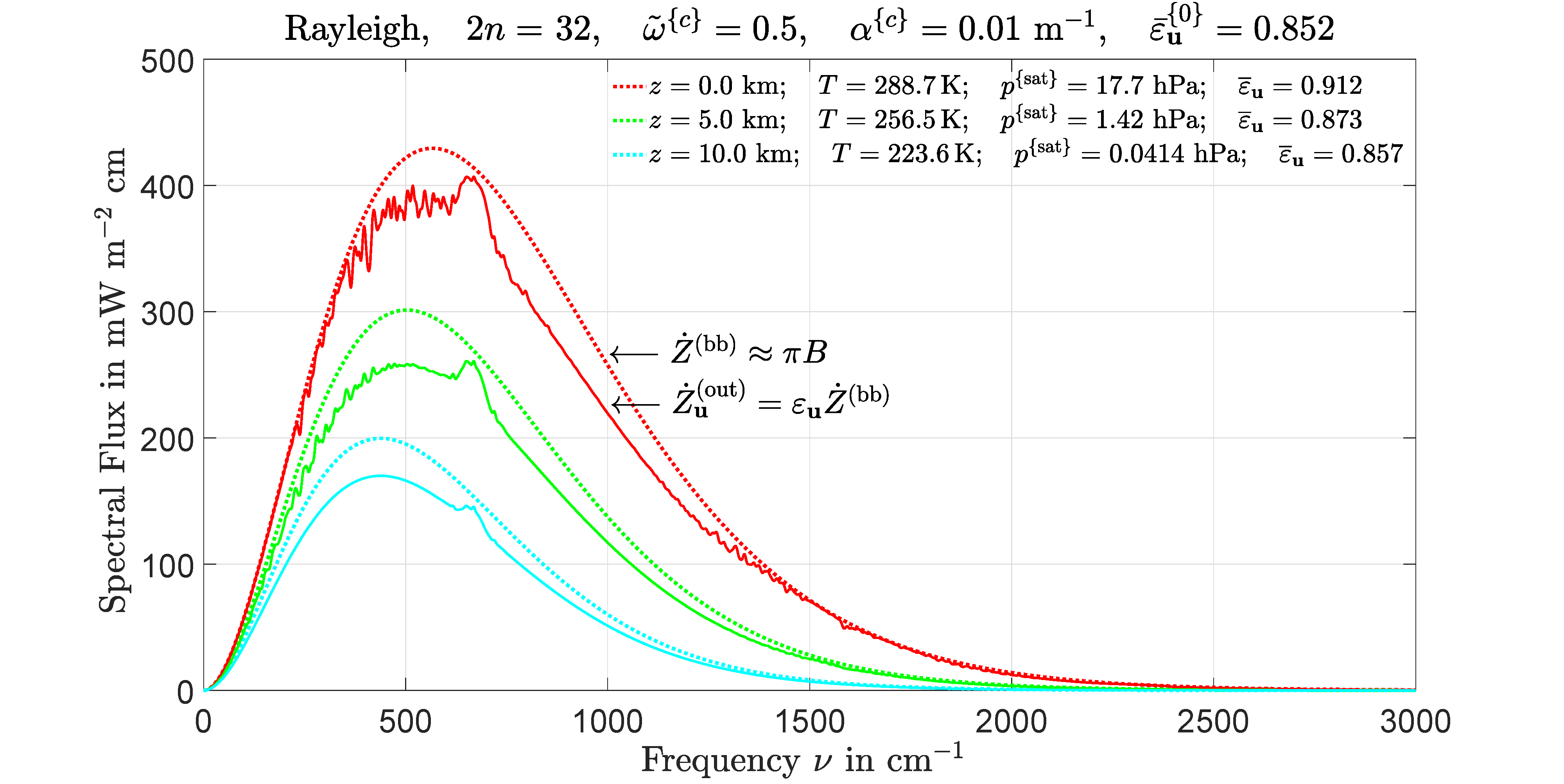}
\caption {Effect of greenhouse gases on the emissivity of optically thick clouds at sea level, $z=0$ km (red), the middle troposphere, $z= 5$ km (green), and near the tropopause $z=10$ km (cyan). A midlatitude standard atmosphere like that discussed in reference\,\cite{WH1} is assumed.
See the text for more discussion.}
\label{emis15}
\end{figure}

Representative spectral fluxes modeled at the tops of optically thick clouds  are shown in Fig. \ref{emis15}. The clouds are located: (1) near the surface (red) where the temperature is $288.7$ K and the saturated vapor pressure of liquid water is $p^{\{\rm sat\}}= 17.7$ hPa; (2) at an altitude $z=5.0$ km (green) where the temperature is 
$256.5$ K and the  saturated vapor pressure of ice is $p^{\{\rm sat\}}= 1.42$ hPa; (3) at an altitude $z=10.0$ km (cyan) where the temperature is 
$223.6$ K and the  saturated vapor pressure of ice is $p^{\{\rm sat\}}= 0.0414$ hPa. As in Fig. \ref{emis3}, the particulates are assumed to have a frequency-independent single-scattering albedo  $\tilde\omega^{\{c\}} =0.5$,  and a frequency-independent attenuation coefficient $\alpha^{\{c\}}=0.01$ m$^{-1}$. The particulates are also assumed to have a Rayleigh scattering phase operator $\hat p$ of (\ref{ls6}), with the multipole coefficients $p_l$ of Table \ref{pl}.

The  blackbody fluxes of (\ref{fe6}), $\dot Z^{(\rm bb)}=4\pi\lvec 0|\hat\mu_{\bf u}|0)B \approx \pi B$, for the various altitudes are shown as dashed lines of Fig. \ref{emis15}. The Planck intensity $B$ was calculated with  (\ref{rtc4}). The upward fluxes of (\ref{fe8}), $\dot Z^{(\rm out)}_{\bf u}=\varepsilon_{\bf u}\dot Z^{(\rm bb)}$, from the tops of optically thick, isothermal clouds containing greenhouse gases are shown as the continuous, somewhat jagged lines in Fig. \ref{emis15}. The values of   $\varepsilon_{\bf u}$ were calculated with (\ref{hm8}). To facilitate graphing of the fluxes $\dot Z^{(\rm out)}_{\bf u}$, the extremely rapid fluctuations with frequency $\nu$  have been smoothed with an area-conserving filter with a width $\Delta\nu = 3$ cm$^{-1}$, as described in Section {\bf 5} of reference\,\cite{WH1}. 

\begin{figure}[t]
\includegraphics[height=100mm,width=1\columnwidth]{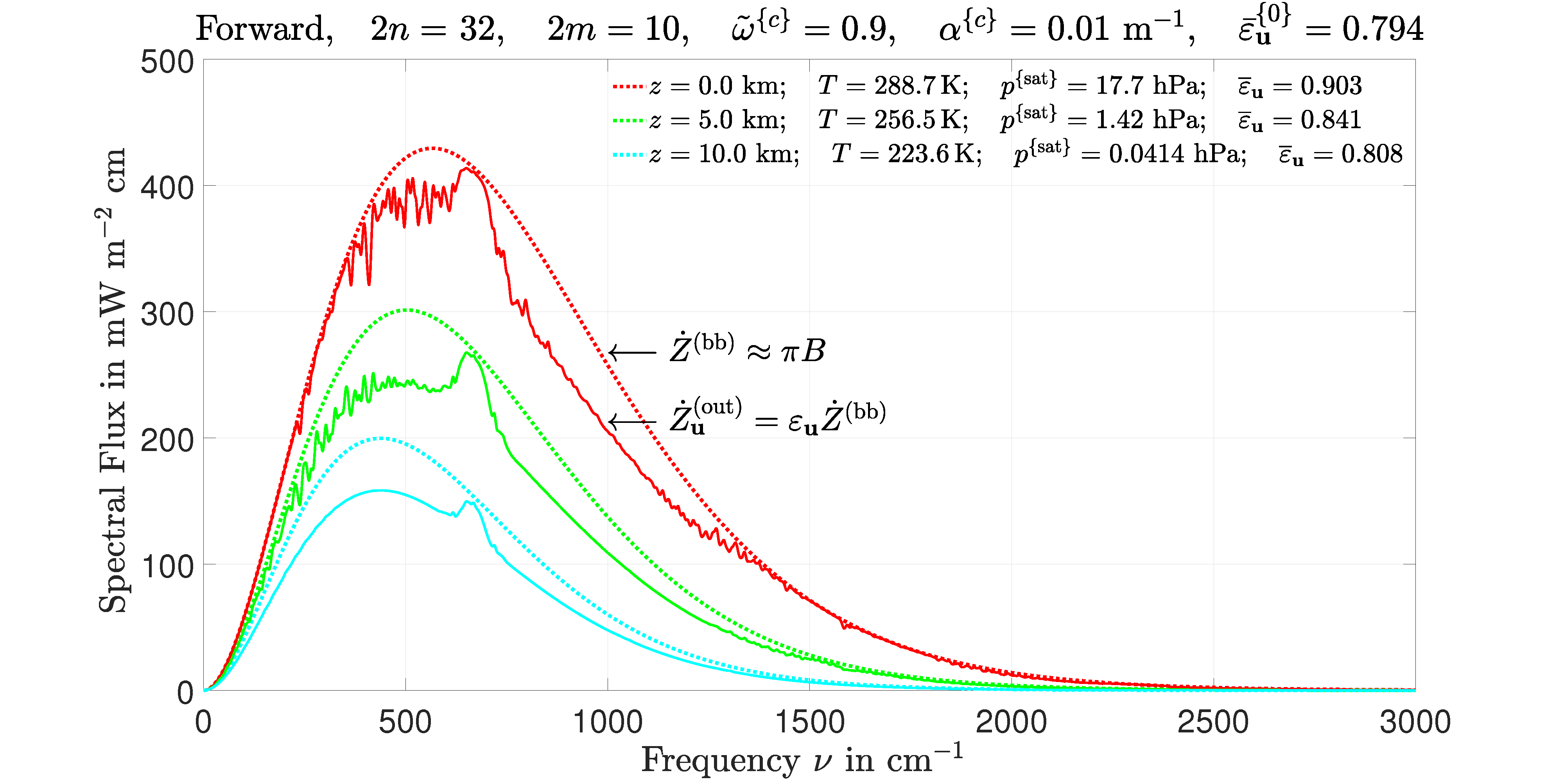}
\caption {Like Fig \ref{emis15} but with forward scattering rather than Rayleigh scattering and with a single-scattering albedo $\tilde\omega^{\{c\}} = 0.9$ rather than $\tilde \omega^{\{c\}} = 0.5$.  See the text for more discussion.}
\label{emis16}
\end{figure}

If the clouds of Fig. \ref{emis15} had no greenhouse gases at all, the frequency-averaged emissivity of (\ref{fie12})  would be $\overline \varepsilon_{\bf u}=\bar\varepsilon_{\bf u}^{\{0\}}=0.852$, as listed in the title of Fig. \ref{emis15}, and as one can estimate from Fig. \ref{emis2}.  For the cloud (cyan), at the highest altitude $z=10$ km, where there is minimal water vapor at the cold temperature, $T=224$ K, the frequency-averaged upward emissivity $\overline \varepsilon_{\bf u}=0.857$ is less than 1\% larger than the limit for no greenhouse gases. The small emissivity enhancement is from the well-mixed greenhouse gas CO$_2$, which is assumed to have a concentration of 400 ppm at all altitudes 

  For a cloud near the surface (red), the partial pressure of water vapor, $17.7$ hPa at a temperature of $T=288.7$ K, is nearly 2\% of the total air pressure, and the frequency-averaged emissivity of  (\ref{fie12}), $\overline \varepsilon_{\bf u}=0.912$, is about 7\% larger than $\bar\varepsilon_{\bf u}^{\{0\}}=0.852$, the emissivity for a cloud with no greenhouse gases at all.

Fig. \ref{emis16} shows what happens when the cloud of of Fig. \ref{emis15} is replaced by one with the forward-scattering phase function shown in Fig. \ref{phase1} or Fig. \ref{phase2} and by a larger single-scattering albedo, $\tilde \omega^{\{c\}}= 0.9$ instead of $\tilde \omega^{\{c\}}=0.5$ as for the cloud of Fig. \ref{emis15}. In this case, the greenhouse-gas-free emissivity of the cloud $\bar\varepsilon_{\bf u}^{\{0\}}= 0.794$ increases by 14\% to $\bar\varepsilon_{\bf u} =0.903$ for the surface cloud.

\section{Summary}
Water vapor, H$_2$O, carbon  dioxide, CO$_2$, and other greenhouse gases within clouds of water droplets or ice crystallites increase the  emissivity and decrease the albedo of the clouds for thermal radiation.
As discussed in Section {\bf \ref{rtp}}, the attenuation coefficient $\alpha$, the single-scattering albedo $\tilde\omega$, and the scattering phase function $p(\mu,\mu')$  are the key parameters of the equation of radiative transfer (\ref{rtc2}). These are determined by the combined effects of cloud particulates and greenhouse gases.  Both cloud particulates and greenhouse gases absorb thermal radiation, but only particulates have appreciable scattering.  Fig. \ref{emis3} shows a representative example of the complicated dependence of the  greenhouse-gas attenuation coefficient $\alpha^{\{g\}}$ of (\ref{df6}) and the single scattering albedo $\tilde\omega$ of (\ref{df12}) on the radiation frequency $\nu$ in the frequency interval from 500 to 700 cm$^{-1}$ which contains many absorption lines of both H$_2$O and CO$_2$ molecules. The complex spectrum is due to the huge numbers of vibration-rotation lines of both molecules.
For frequencies at the peaks of the absorption coefficient $\alpha^{\{g\}}$, where cloud particulates make negligible contributions to the total  absorption coefficient $\alpha$, the single scattering albedo  nearly vanishes, $\tilde\omega\to 0$, an isothermal cloud will appear black and it will have a Lambertian (uniform) angular distribution of near unit intensity emissivities, $\varepsilon_{k}\approx 1$. For frequencies  in atmospheric windows where the greenhouse-gas absorption coefficient is small, $\alpha^{\{g\}}\ll\alpha^{\{c\}}$, the intensity emissivities $\varepsilon_k$ will be smaller than unity and the angular dependence of the emissivities will be limb darkened.  The effects of greenhouse gases on cloud emissivity are most pronounced for frequencies where  the attenuation coefficient $\alpha^{\{g\}}$ of the greenhouse gases is large compared to the attenuation coefficient $\alpha^{\{c\}}$ of cloud particulates,  that is, when $\alpha^{\{g\}}>\alpha^{\{c\}}$.

In Section {\bf\ref{int}}, we review mathematical methods that can be used to solve the $2n$-stream equation of transfer (\ref{rtc6}). Axially symmetric radiation, propagating with direction cosine $\mu$ with respect to the vertical, can be represented by the intensity function $I(\mu)$ which has an infinite number of values, one for each real number $\mu$ in the interval $-1\le\mu\le 1$.  But for most practical problems, this is far more information than needed to accurately calculate atmospheric heating or cooling from greenhouse gases.  In $2n$-stream  radiation transfer theory, the continuous function $I(\mu)$ is approximated by  $2n$ sample values $I(\mu_i)$ at the Gauss-Legendre direction cosines $\mu_i$, which are the zeros of the Legendre polynomial $P_{2n}$ and  defined by $P_{2n}(\mu_i)=0$. This allows one to replace the exact integro-differential equation (\ref{rtc2}) by the ordinary vector differential equation (\ref{rtc6}),  which is much easier to solve than (\ref{rtc2}) and which gives radiative heating and cooling rates that are practically indistinguishable from the exact solutions.

In section {\bf\ref{cl}} we show how to analyze the effects of greenhouse gases on the thermal emission of clouds.  The fundamental  tool for calculations is the scattering operator 
$\mathcal{S}$ and its complement, the emissivity operator $\mathcal{E}=\hat 1-\mathcal{S}$, which are discussed in Section {\bf\ref{sm}}.
Emissivities for isothermal clouds are quantified by the intensity emissivities $\varepsilon_i$ discussed in Section {\bf\ref{ie}}, and by the upward and downward flux emissivities, 
$\varepsilon_{\bf u}$  and $\varepsilon_{\bf d}$  discussed in Section {\bf\ref{fe}}.  As shown in Section  {\bf\ref{be}}, the emissivities are nonnegative and no greater than 1. The emissivities are particularly easy to calculate for homogeneous clouds, which are discussed in Section {\bf\ref{hc}}.  

Examples of intensity emissivities are shown in Fig. \ref{aemis2} for Rayleigh-scattering cloud particulates. Fig. \ref{aemis3} shows similar examples  for strongly forward-scattering particulates. Black clouds, with vanishing single scattering albedos, $\tilde \omega=0$, have unit emissivities, $\varepsilon_i = 1$, and Lambertian angular dependences. As the single-scattering albedos $\tilde \omega$ of particulates in isothermal, optically-thick clouds increase, the intensity emissivities $\varepsilon_k$ decrease, especially for slant streams with $|\mu_k|\ll 1$. This causes the  limb darkening of the emissivities of Fig. \ref{aemis2} and Fig. \ref{aemis3} with increasing $\tilde\omega$.

Fig. \ref{emis2} shows how the upward flux emissivities $\varepsilon_{\bf u}$ of optically thick, homogeneous clouds depend on the single-scattering albedo $\tilde\omega$ for each of the four scattering phase functions $p(\mu)$ of (\ref{spf10}) shown in Fig. \ref{phase1} and Fig. \ref{phase2}. The phase functions are parameterized by the multipole coefficients $p_l$ of Table~\ref{pl}. Increasing single scattering albedos $\tilde \omega$  decrease the upward flux emissivities $\varepsilon_{\bf u}$. For all phase functions,  $\varepsilon_{\bf u}\to 1$ as $\tilde\omega\to 0$ and $\varepsilon_{\bf u}\to 0$ as $\tilde\omega\to 1$. For a given single-scattering albedo $\tilde\omega$, clouds with forward scattering have the largest emissivities and clouds with backward scattering have the smallest.

Fig. \ref{emis15} shows how  the greenhouse gases H$_2$O and CO$_2$ modify the emissivity of clouds with  Rayleigh scattering phase functions, frequency-independent single-scattering albedos $\tilde\omega^{\{c\}}=0.5$ and frequency-independent attenuation coefficients $\alpha^{\{c\}}=0.01$ m$^{-1}$. Fig. \ref{emis16} shows what happens if the Rayleigh scattering phase function used for Fig. \ref{emis15} is replaced by a forward scattering phase function, and if the particulates are assigned a larger single-scattering albedo, $\tilde\omega^{\{c\}}=0.9$ instead of $\tilde\omega^{\{c\}}=0.5$.

It is straightforward to model other combinations of the basic cloud parameters, $\alpha^{\{c\}}$, $\omega^{\{c\}}$ and $p^{\{c\}}(\mu,\mu')$. But we have limited our discussion to a few examples to give a quick,  semiquantitative overview of the main facts. Typical modifications of cloud emissivities and albedos by greenhouse gases are on the order of 10\%. The modifications are least at high, cold altitudes, where the saturation vapor pressure of water is relatively small, and greater near the surface where the vapor pressure is much larger. We have only considered H$_2$O line structure, and including opacity of the water vapor continuum  would  further increase the emissivity of low-altitude clouds\,\cite{continuum}. A 10\%  effect may seem small, but the influence of increasing greenhouse gases is even smaller. ``Instantaneously'' doubling the concentration of CO$_2$ in cloud-free air only decreases the radiation flux to space by about 1\%, as discussed in reference\,\cite{WH1}. The enhancement of  thermal emission by greenhouse gases  in low clouds should therefore be taken into account for accurate calculations of radiation transfer in Earth's atmosphere.

\section*{Acknowledgements}
  The Canadian Natural Science and Engineering Research  Council provided financial support for one of us.

\end{document}